%% file: main.tex
\documentclass[reprint,preprintnumbers,amsmath,amssymb,aps,prx,superscriptaddress,nofootinbib]{revtex4-2}

\usepackage{graphicx}
\usepackage{dcolumn}
\usepackage{bm}
\usepackage{hyperref}
\usepackage[mathlines]{lineno}
\usepackage{xcolor}
\usepackage{booktabs}
\usepackage{enumitem,kantlipsum}
\usepackage{wasysym}
\usepackage{siunitx}
\usepackage{cleveref}
\usepackage{xspace}

\newcommand{\ddjj}{$\delta\Delta_{\mathrm{JJ}}$\xspace}
\newcommand{\ddgnd}{$\delta\Delta_{\mathrm{M1}}$\xspace}
\newcommand{\Am}{$^{241}$Am\xspace}
\newcommand{\fq}{$f_{\mathrm{qb}}$\xspace}
\newcommand{\MG}{Ref.~\onlinecite{marchegianiNonequilibriumRegimesQuasiparticles2025}\xspace}
\DeclareSIUnit[]\electron
{\text{\ensuremath{e^{-}}}}
\DeclareSIUnit[quantity-product = ~]{\mwe}{\text{m.w.e.}}
\newcommand{\g}{$|0\rangle$\xspace}
\newcommand{\e}{$|1\rangle$\xspace}

\begin{document}

\preprint{}


\title{Characterization of Radiation-Induced Errors in Superconducting Qubits Protected with Various Gap-Engineering Strategies}

\include{alpha_linac_authorlist}

\date{\today}

\begin{abstract}

Impacts from high-energy particles cause correlated errors in superconducting qubits by increasing the quasiparticle density in the vicinity of the Josephson junctions (JJs). Such errors are particularly harmful as they cannot be easily remedied via conventional error correcting codes. 
Recent experiments reduced correlated errors by making the difference in superconducting gap energy across the JJ larger than the qubit transition energy. 
In this work, we assess gap engineering near the JJ (\ddjj) and the capacitor/ground-plane (\ddgnd) by exposing arrays of transmon qubits to two sources of radiation.
For $\alpha$-particles from an \Am source, we observe $T_1$ errors correlated in space and time, supporting a hypothesis that hadronic cosmic rays are a major contributor to the $10^{-10}$ error floor observed in Ref.~\onlinecite{acharyaQuantumErrorCorrection2025}.
For electrons from a pulsed linear accelerator, we observe temporally correlated $T_1$ and $T_2$ errors, this measurement is insensitive to spatial correlations.
We observe that the severity of correlated $T_1$ errors is reduced for qubit arrays with a greater degree of gap engineering at the JJ.  
For both $T_1$ and $T_2$ errors, the recovery time is hastened by an increased \ddgnd, which we attribute to the trapping of quasiparticles into the capacitor/ground-plane.   
We construct a model of quasiparticle dynamics that qualitatively agrees with our observations.
This work reinforces the multifaceted influence of radiation on superconducting qubits and provides strategies for improving radiation resilience.

\end{abstract}

\maketitle


\section{Introduction}\label{sec:intro} 

Interactions of environmental radiation
with the substrate of superconducting circuits are a source of non-equilibrium material excitations.  These interactions generate phonons by (1) creating electron-hole pairs that can recombine into phonons and (2) direct nuclear displacement.  In turn, these phonons can interact with superconducting materials to create Bogoliubov quasiparticles (QPs).  This mechanism is widely leveraged as a radiation-detection technique~\cite{irwinQuasiparticletrapassistedTransitionedgeSensor1995, dayBroadbandSuperconductingDetector2003} and has been identified as a source of correlated errors in superconducting qubits~\cite{vepsalainenImpactIonizingRadiation2020, wilenCorrelatedChargeNoise2021, thorbeckTwoLevelSystemDynamicsSuperconducting2023, harringtonSynchronousDetectionCosmic2025, liCosmicrayinducedCorrelatedErrors2025, bratrudMeasurementCorrelatedCharge2025, dominicisEvaluatingRadiationImpact2024,ql6q-wfpn,kurilovichCorrelatedErrorBursts2025,larsonQuasiparticlePoisoningSuperconducting2025}.

The effects of radiation on qubits are correlated in both space and time~\cite{wilenCorrelatedChargeNoise2021,mcewenResolvingCatastrophicError2022, yeltonModelingPhononmediatedQuasiparticle2024,acharyaQuantumErrorCorrection2025,kurilovichCorrelatedErrorBursts2025}, introducing qubit frequency shifts and state transition errors that are difficult to rectify within standard quantum error correction (QEC) frameworks~\cite{tanResilienceSurfaceCode2024,acharyaQuantumErrorCorrection2025,kurilovichCorrelatedErrorBursts2025}.  While QEC schemes are being developed to accommodate correlated errors~\cite{suzukiQ3DEFaulttolerantQuantum2022,strikisQuantumComputingScalable2023, tanResilienceSurfaceCode2024,valleroEfficacySurfaceCodes2024a}, they come at the cost of increased resource requirements.  Device-level mitigation strategies could reduce the occurrence, extent, and severity of such correlated errors --- and thereby the associated resource overhead.

One strategy for device-level mitigation of relaxation errors is by engineering the superconducting gap in the vicinity of the Josephson junction (JJ)~\cite{aumentadoNonequilibriumQuasiparticles$2e$2004, mcewenResistingHighEnergyImpact2024, ql6q-wfpn,kurilovichCorrelatedErrorBursts2025,harringtonAsymmetricJosephsonJunctions2025}.  In qubits with Al/AlO$_\text{x}$/Al JJs this can be achieved by tuning the thickness of the Al layers comprising the JJ~\cite{marchegianiQuasiparticlesSuperconductingQubits2022}.  If the gap difference across the JJ (\ddjj) is greater than the qubit energy ($h$\fq), QPs thermalized to the gap edge will not be able to facilitate relaxation errors by absorbing a qubit photon and tunneling in one direction.  We refer to this as ``JJ gap-engineering'' (JJ GE).  Previous work has used this technique to make qubits insensitive to the higher-rate, lower-energy (below \qty{1}{\MeV} deposited) portion of the natural radiation spectrum~\cite{mcewenResistingHighEnergyImpact2024}, which consists primarily of $\gamma$-rays from trace radioactive elements and cosmic ray leptons (electrons and muons)~\cite{fowlerSpectroscopicMeasurementsModels2024, loerAbatementIonizingRadiation2024}.

Despite the progress of JJ GE and other mitigation strategies~\cite{panEngineeringSuperconductingQubits2022,iaiaPhononDownconversionSuppress2022,bargerbosMitigationQuasiparticleLoss2023,yeltonModelingPhononmediatedQuasiparticle2024,kamenovSuppressionQuasiparticlePoisoning2024,larsonQuasiparticlePoisoningSuperconducting2025,wuMitigatingCosmicraylikeCorrelated2025,bertoldoCosmicMuonFlux2025,kurilovichCorrelatedErrorBursts2025,ql6q-wfpn}, a residual source of correlated errors (approximately one-per-hour) apparently remains a limiting factor in state-of-the-art, below-threshold QEC~\cite{acharyaSuppressingQuantumErrors2023, acharyaQuantumErrorCorrection2025,kurilovichCorrelatedErrorBursts2025}.  This rate is consistent with what would be expected from the higher-energy (above \qty{1}{\MeV} deposited) portion of the natural radiation spectrum, which consists primarily of cosmic-ray hadrons (protons and neutrons)~\cite{fowlerSpectroscopicMeasurementsModels2024, loerAbatementIonizingRadiation2024}.  These hadronic cosmic rays could be expected to overcome the protection of JJ GE due to the order-of-magnitude greater energy deposited in the qubit substrate.  Potential sensitivity to these energies, which can also be reached by terrestrial $\alpha$-particles impinging on the chip, would not have been readily discerned in previous work due to the short duration (\SI{100}{\minute}) of the experiment~\cite{mcewenResistingHighEnergyImpact2024}.  

Our work has three main goals: (1) to investigate whether these hadronic comic rays can explain the residual correlated error rate observed~\cite{acharyaQuantumErrorCorrection2025}, (2) to study the potential limits of JJ GE, and (3) to quantify the effect of a superconducting gap difference between the metal forming the JJ and the metal forming the capacitor/ground-plane.  We refer to the superconducting gap difference at this interface as \ddgnd.  To this end we designed and fabricated arrays of transmon qubits with three superconducting gap profiles and exposed them to two distinct types of radiation: \qty{5.5}{\MeV} $\alpha$-particles from \Am and \qtyrange{10}{20}{\mega\electronvolt} electrons ($e^-$) from a linear accelerator (linac) at the Johns Hopkins Applied Physics Laboratory (APL) Controlled Linac Irradiation of Quantum Experiments (CLIQUE) facility.  

The measurements with \Am described in Sec.~\ref{sec:am_source} serve two purposes.  Primarily they address goal 1, informing whether low-rate, high-energy cosmic ray hadrons can be a source of the correlated errors observed in recent QEC experiments~\cite{fowlerSpectroscopicMeasurementsModels2024,acharyaQuantumErrorCorrection2025}.  In addition, they directly enable us to study the effects of $\alpha$-particle impacts on qubit substrates.  These particles can be emitted by radioactive contaminants in materials that form the qubit package (e.g., ceramic interposer PCBs, connectors, and dust)~\cite{fowlerSpectroscopicMeasurementsModels2024, loerAbatementIonizingRadiation2024, cardaniDisentanglingSourcesIonizing2023,akeribLUXZEPLINLZRadioactivity2020} or by radon emanated from qubit package materials~\cite{akeribLUXZEPLINLZRadioactivity2020}, and are a well known and challenging background to mitigate in rare-event particle physics experiments~\cite{thelux-zeplincollaborationBackgroundDeterminationLUXZEPLIN2023}.  

Our measurements at CLIQUE, presented in Sec.~\ref{sec:linac}, allow us to precisely investigate the influence of the superconducting gap profile on radiation response, addressing goals 2 and 3.  These measurements leverage the facility's two key advantages: triggered timing and tunable beam current.  The timing and \SI{10}{\hertz} repetition rate of the electron beam trigger enabled the collection of datasets with $\mathcal{O}(1000)$ samples of $e^-$-qubit interactions at a range of energies.  These individual interactions are averaged to reduce sensitivity to non-radiation effects.  We probed the energy dependence of qubit response by tuning the beam current of the linac.  

Throughout this work we measure radiation's effect on qubit excitation and relaxation rates.  Additionally, we employ pulse sequences tailored to detect transient qubit frequency shifts~\cite{kurilovichCorrelatedErrorBursts2025,mcjunkinOnDemandCorrelatedErrors2026}.  We find that each quantity's response to radiation is influenced, but not eliminated, by the qubit's superconducting gap profile.  We conclude by constructing a model connecting QP dynamics with qubit transition rates, and demonstrate that this model re-creates our key observations, Sec.~\ref{sec:modeling}.


\section{Qubit Arrays and Gap Profiles Under Study}

\begin{figure*}[htbp]
    \begin{center}
        \includegraphics[width=0.95\textwidth]{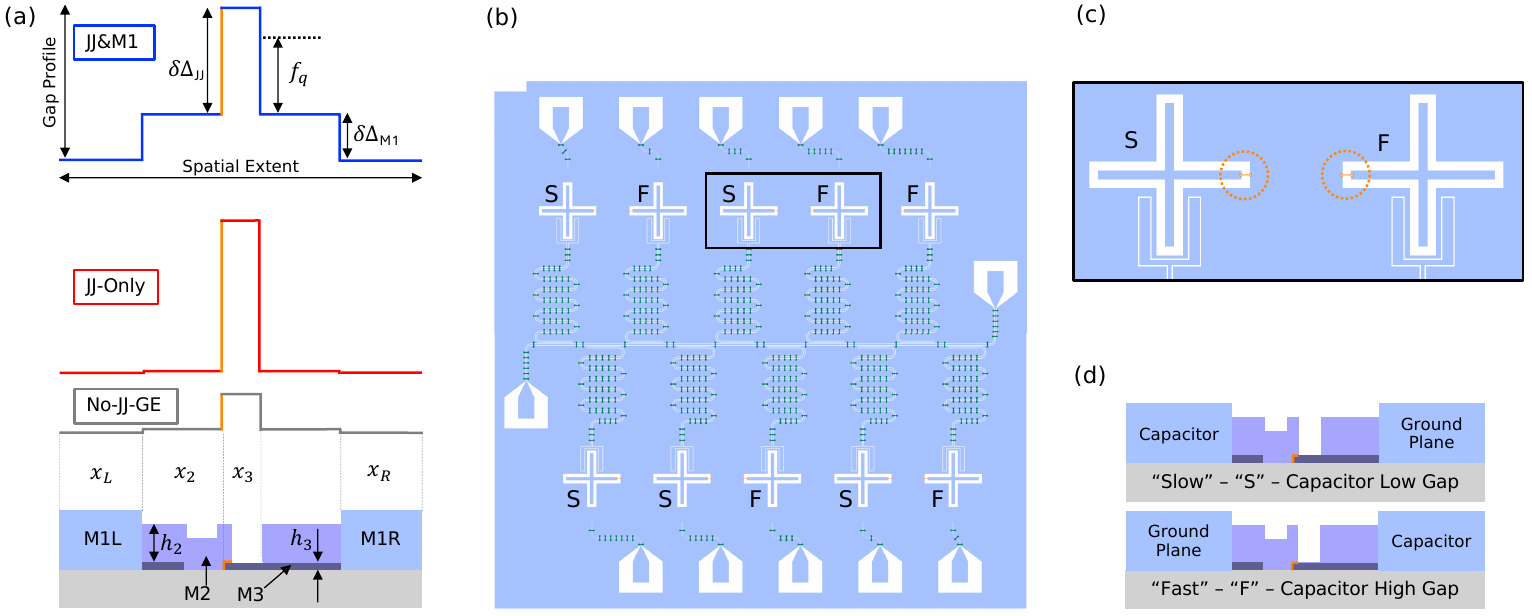}
    \end{center}
    \caption{Qubit arrays under study.  (a) The three superconducting gap profiles investigated above a cartoon cross section of a Dolan bridge JJ.  Vertical distance is proportional to $\Delta - \Delta_\text{M1}$.  Axes and the parameters \ddjj, \fq, and \ddgnd are defined for the JJ\&M1 array, with further information in Table~\ref{tab:delta_deltas}.  For gap profiles, the location of the JJ is shown as an orange vertical line.  The bottom row shows a cartoon cross-section of a Dolan bridge JJ.  Metal layers are labeled and ordered by their superconducting gaps: M1L and M1R (light blue) are the etched base metal on the left and right sides of the JJ and constitute the lowest gap layer on the chip, M2 (light purple) is the lower gap (thicker) shadow evaporated layer, and M3 (dark purple) is the higher gap (thinner) shadow evaporated layer.  The JJ is shown in orange.  The QP densities modeled in Section~\ref{sec:model_main_text} are indicated as $x_i$, where the subscript $i$ indicates the relevant metal, shortened to $L$, $R$, $2$, and $3$ for brevity.  (b) An artificially colored layout of the qubit array geometry.  Layer M1 is light blue, air-bridges are green, and the JJ region is orange.  The black box highlights the expanded view in panel (c).  Qubit JJ orientation is indicated in black text.  (c) Expanded view of two qubits in the array, highlighting the slow (S) and fast (F) JJ orientations.  Layer M1 is light blue while the JJ is circled and highlighted in orange.  (d) Cartoon cross-sections of Dolan bridge JJs highlighting the difference between JJ orientations S and F.}
    \label{fig:devices}
\end{figure*}

\begin{table*}[htbp]
    \centering
    \begin{tabular}{l l l l l l}
        \toprule
        Array           & \hspace{0.5cm} $\bar{f}_\text{qb}$~(GHz)  &  \hspace{0.5cm} \ddjj/$h$~(GHz) &  \hspace{0.5cm} \ddgnd/$h$~(GHz)  &  \hspace{0.5cm} h$_2$~(nm)  &  \hspace{0.5cm} h$_3$~(nm)  \\
        \midrule
        No-JJ-GE  & \hspace{0.5cm} 4.210 &  \hspace{0.5cm} 2.8         &  \hspace{0.5cm} 0.2           &  \hspace{0.5cm} 160         &  \hspace{0.5cm} 30          \\
        JJ-Only & \hspace{0.5cm} 5.252 &  \hspace{0.5cm} 10.0        &  \hspace{0.5cm} 0.1           &  \hspace{0.5cm} 200         &  \hspace{0.5cm} 10          \\
        JJ\&M1 & \hspace{0.5cm} 5.207 &  \hspace{0.5cm} 7.0         &  \hspace{0.5cm} 3.0           &  \hspace{0.5cm} 30          &  \hspace{0.5cm} 10          \\
        \bottomrule
    \end{tabular}
    \caption{Gap engineering parameters for the three arrays under test.  Average qubit frequencies $\bar{f}_q$ are measured, while \ddjj, \ddgnd, h$_2$, and h$_3$ are the nominal designed values, informed by transition temperature measurements of films from the same fabrication process.}
    \label{tab:delta_deltas}
\end{table*}

This work focuses on three chips, each with ten individual transmons and nearly identical geometric designs.  Each chip has a distinct superconducting gap profile, sketched to scale in Fig.~\ref{fig:devices}(a).  All layout features are identical to the array described in Ref.~\cite{harringtonSynchronousDetectionCosmic2025} with the exception of JJ lead thicknesses as described in Table~\ref{tab:delta_deltas} and Fig.~\ref{fig:devices}.  We refer to the three metal layers as M1, M2, and M3, ordered from lowest to highest superconducting gap [Fig.~\ref{fig:devices}(a)].  Layers M1L and M1R are the base metalization and form the ground plane and qubit capacitor, with L and R indicating position relative to the JJ.  Layers M2 and M3 form the JJ.  The substrate of each array was a $\qty{5}{\mm} \times \qty{5}{\mm} \times \qty{350}{\um}$ silicon chip.

The JJs on qubits in these arrays are distributed between two orientations: ``slow,'' (S) where the low gap side of the JJ contacts the relatively small, isolated qubit capacitor, and ``fast,'' (F) where the low gap side of the JJ contacts the relatively large ground plane [Fig.~\ref{fig:devices}(d)].  In Ref.~\onlinecite{harringtonSynchronousDetectionCosmic2025}, this difference was shown to influence the qubit relaxation error recovery timescale.  It was hypothesized that since relaxation errors are driven by the QP population in the relatively low gap M2 layer, the QP dynamics differ between the qubit capacitor and the ground plane.  This was attributed to a difference in trapping efficiency, caused by the ground plane containing more features which can remove QPs from the system (e.g., lower-gap air bridges and galvanic contact to the normal metal of the qubit package).  In this work, we use the difference in response between fast and slow JJ orientations to identify the influence of QP dynamics in layer M1 on the qubits.

The three superconducting gap profiles studied are as follows.  There are three relevant energy scales: the qubit energy $h$\fq and the two superconducting gap differences \ddjj$=\Delta_\text{M3} - \Delta_\text{M2}$ and \ddgnd$=\Delta_\text{M2} - \Delta_\text{M1}$.  For fabrication consistency, the thinnest M3 layer used was \qty{10}{\nano\meter}, limiting the sum $(\delta\Delta_\text{JJ} + \delta\Delta_\text{M1})/h$ to be less than approximately \qty{10}{\giga\hertz}.  The \underline{No-JJ-GE} array acts as a control and was fabricated at the same time as the chip studied in Ref.~\onlinecite{harringtonSynchronousDetectionCosmic2025}.  There was no special care taken to mitigate correlated errors from environmental radiation, and so \ddjj was less than the qubit energy, and \ddgnd/$h$ had a designed value of \qty{0.2}{\giga\hertz}.  The next array, \underline{JJ-Only}, had the largest designed \ddjj/$h$ at \qty{10}{\giga\hertz} and the lowest designed \ddgnd/$h$ at \qty{0.1}{\giga\hertz}.  This profile is in the style of those measured in Refs.~\onlinecite{mcewenResistingHighEnergyImpact2024,acharyaQuantumErrorCorrection2025,kurilovichCorrelatedErrorBursts2025}.  The final array, \underline{JJ\&M1}, also has a nominal \ddjj greater than the qubit energy.  Additionally, a nominal \ddgnd/$h$ of approximately \SI{3}{\giga\hertz} was introduced to test the efficacy of using layer M1 as a QP trap.  This large \ddgnd limits the magnitude of \ddjj on this array to a designed value of \qty{7}{\giga\hertz}.  Table~\ref{tab:delta_deltas} contains the nominal gap engineering parameters, estimated from transition temperature measurements on films from the same fabrication process, as well as measured average qubit frequencies.  

As the no-JJ-GE array was thoroughly characterized in Refs.~\onlinecite{harringtonSynchronousDetectionCosmic2025,mcjunkinOnDemandCorrelatedErrors2026}, we focus our discussion on the two arrays designed with \ddjj~$>h$\fq.  Typical energy relaxation times $T_1$ for the qubits under study were greater than \SI{25}{\micro\second} (see Table~\ref{tab:t1_and_freq} in Appendix~\ref{appendix:linac} for further details).  Throughout, we refer to the qubit ground state as \g and the $i^\text{th}$ excited state as $|i\rangle$.  Unless otherwise stated, time refers to the time since a particle-qubit interaction.


\section{Qubit response to $\alpha$-particles}
\label{sec:am_source}

\subsection{Measurement Setup}\label{alpha_setup}

Sensitivity to high-energy particles, such as $\alpha$-particles and hadronic cosmic rays, was tested using a \SI{37}{\kilo\becquerel} \Am source inside of a dilution refrigerator at MIT.  This source emits both \SI{5.5}{\MeV} $\alpha$-particles and \SI{59}{\keV} $\gamma$-rays. 
The source was glued to a copper mount approximately \SI{3}{\cm} away from the qubit substrate surface and faced the side with the qubits.  When measuring the JJ-Only array, the source was collimated using an approximately \SI{100}{\micro\m} diameter pinhole in copper tape, and when measuring the JJ\&M1 array, the source was collimated using an \qty{7.96}{\milli\meter} thick custom machined block of Cu101 with three \qty{180}{\um} diameter holes.  Both collimators were designed such that the $\alpha$-particle impact rate was approximately~\qty{0.077}{\per\s}. An illustration of this setup is shown in Fig.~\ref{fig:am-241} of Appendix~\ref{appendix:am-241}.  The data acquisition is similar to what is described in Refs.~\onlinecite{harringtonSynchronousDetectionCosmic2025,vepsalainenImpactIonizingRadiation2020}.  

\subsection{Data Collection and Analysis}\label{sec:am_data_and_analysis}

\begin{figure}[tbp]
    \begin{center}
    \includegraphics[width=\columnwidth]{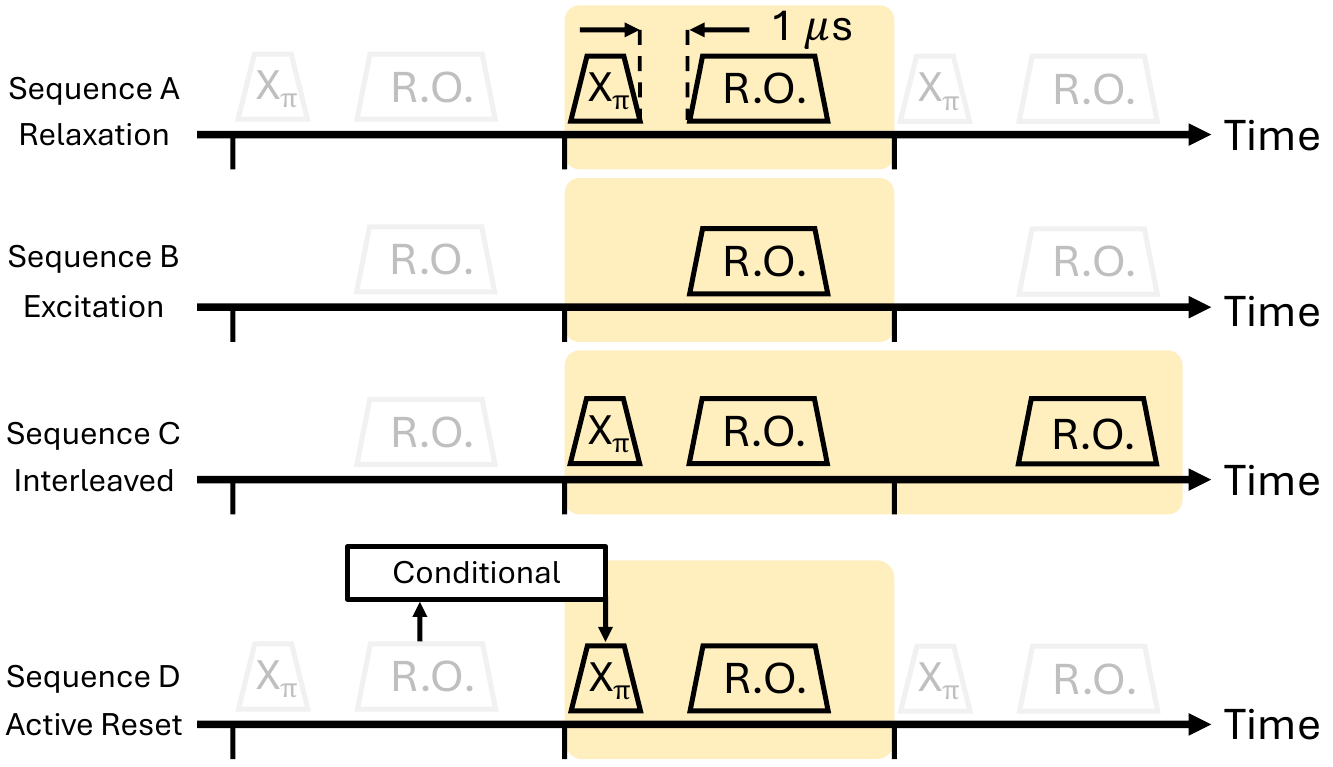}
    \end{center}
    \caption{Measurement sequences used for identifying changes in qubit transition rates.  Sequences A-C are heralded measurements used during the collection of data with the \Am source.  Sequence D is an active reset measurement used at CLIQUE.  With sequence D, sensitivity to qubit relaxations (excitations) was improved relative to sequences A and B by conditionally applying X$_\pi$ to increase state preparation in \e (\g).}
    \label{fig:sequences}
\end{figure}

Three datasets were collected with each array, each using a unique measurement sequence [Fig.~\ref{fig:sequences}]: (A) a measurement primarily sensitive to qubit relaxation, (B) a measurement primarily sensitive to qubit excitation, and (C) a measurement sequence that alternates between these two. Appendix~\ref{sec:appendix_measurement_sequences} contains additional details about the measurement sequences.  Sequence A comprised a repeated sequence of an X$_\pi$ (a rotation of $\pi$~radians around the X-axis of the Bloch sphere), a \SI{1}{\micro\s} delay, and a \SI{4}{\micro\s} readout.  The measurement uses ``heralding'', meaning the previous cycle's measurement was used to determine whether a qubit relaxation occurred, defined to be whenever the qubit was measured in \g twice in a row:

\begin{align}\label{eqn:relaxation_sequence}
    \text{Measured~in~\g~}\rightarrow~\mathrm{X}_\pi\text{~to~\e~}\rightarrow\text{~Measure in~\g}.
\end{align}

\noindent Sequence B is the same as sequence A, except the X$_\pi$ is omitted.  In this sequence, an excitation is measured whenever \e follows \g:

\begin{align}\label{eqn:excitation_sequence}
    \text{Measured~in~\g~}\rightarrow~\text{no~}\mathrm{X}_\pi\rightarrow\text{~Measure in~\e}.
\end{align}

\noindent Sequence C alternates whether the X$_\pi$ is applied, allowing for identification of qubit relaxations on measurements bracketing an X$_\pi$ [condition~\ref{eqn:relaxation_sequence}] and excitation on measurements not bracketing an X$_\pi$ [condition~\ref{eqn:excitation_sequence}].  

An aspect of these sequences is that the qubit trajectory is ambiguous following an excited-state measurement.  For example, in the qubit relaxation sequence (A), measuring \e~$\rightarrow$~\g could be due to two relaxation errors (relaxation both before and after the X$_\pi$), or no relaxation errors (the X$_\pi$ caused the state transition).  We remove such cases from our analysis.

Each dataset comprised \num{250e6} measurement cycles.  The cycle duration was \SI{6.95}{\micro\s}, resulting in a dataset with approximately \qty{29}{\minute} of live time.  Increases in qubit relaxation rates due to radiation were identified using the same matched filter procedure as was used in Ref.~\onlinecite{harringtonSynchronousDetectionCosmic2025} (see Supplementary Note 1 therein) except using a \qty{100}{\micro\s} decaying exponential as the filter template (cf., a \qty{5}{ms} decaying exponential that integrated to 0).  As the change in qubit excitation rate due to radiation is a smaller signal relative to the relaxation rate, identifying them with sequence B (primarily sensitive to qubit excitations) was inefficient.  To mitigate this type of signal-to-noise ratio issue, we instead identified qubit excitations using sequence C: a matched filter is used to identify radiation impacts from increases in the qubit relaxation rate with measurements that bracket an X$_\pi$, and qubit excitations for those same radiation impacts are studied with measurements that do not bracket an X$_\pi$.  Details on identifying radiation impacts are presented in Appendix~\ref{appendix:alpha_data_selection}.

\subsection{Results and Interpretation}

\begin{figure}[tbp]
    \begin{center}
        \includegraphics[width=\columnwidth]{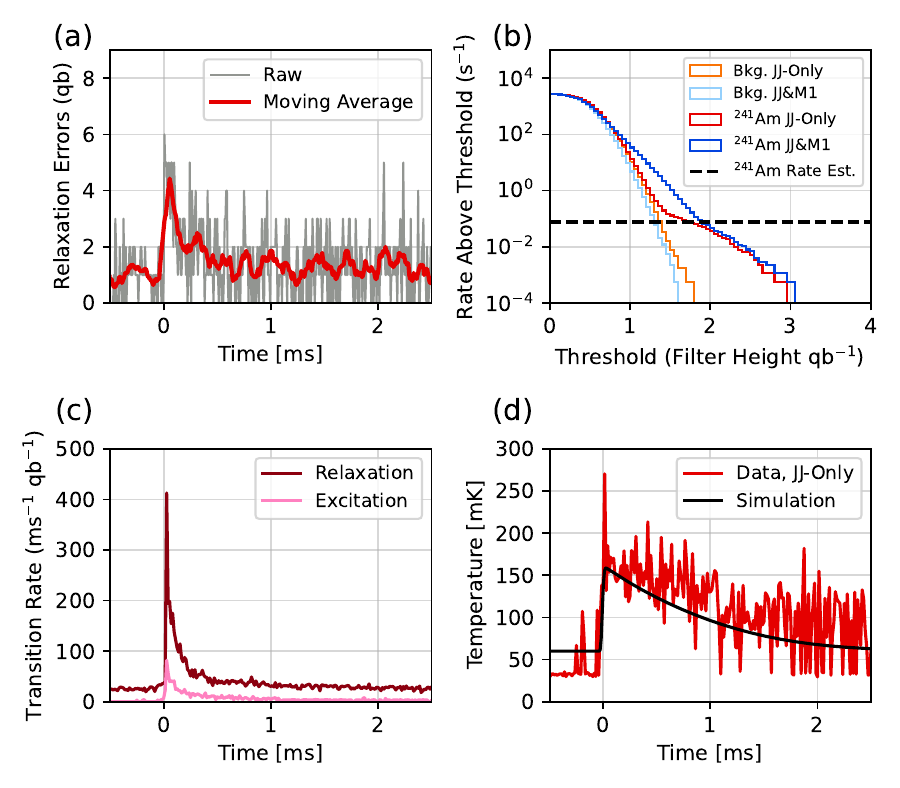}
    \end{center}
    \caption{Qubit response from the \Am source.  (a) The largest particle interaction recorded for the JJ-Only array.  Data were collected using sequence A.  Raw data are gray while a 14-point moving average is shown in red as a guide to the eye.  The vertical axis shows the total number of qubit errors in the array per measurement cycle.  (b) The detected correlated error rate integrated above various values for background datasets (orange, light blue) and \Am source datasets (red, blue).  Warm colors (orange and red) represent the JJ-Only array, cool colors (light blue and blue) represent the JJ\&M1 array.  A black dashed line indicates the event rate expected from the geometry of our collimation.  Data were collected using sequence A. (c) Median qubit relaxation (red) and excitation (pink) transition rate per qubit resulting predominantly from $\alpha$-particle impacts. Data were collected using sequence C. (d) The estimated median qubit temperature following an $\alpha$-particle impact is shown in red.  A simulated thermal response from an $\alpha$-particle impact event is shown in black.  Details of the simulation are presented in Section~\ref{sec:model_comparison}.}
    \label{fig:alphas}
\end{figure}

We display results from these datasets in Fig.~\ref{fig:alphas}.  Panels (a) and (b) use sequence A, while (c) and (d) use sequence C.  The JJ-Only array is shown in all panels while the JJ\&M1 array is only shown in Fig.~\ref{fig:alphas}(b) --- the fast recovery timescale of the JJ\&M1 array resulted in a lower trigger efficiency for sequence C relative to sequence A.

Despite strong protection at the JJ, \ddjj~$>$~$h$\fq, we observe an increase in the rate of correlated qubit errors with respect to background in both arrays [Fig.~\ref{fig:alphas}(a, b)].  These correlated errors are created by transient increases in both the qubit relaxation \textit{and} excitation rates [Fig.~\ref{fig:alphas}(c)]. The characteristic recovery timescale of these increased transition rates is approximately \qty{100}{\us}, an order of magnitude faster than the No-JJ-GE array~\cite{harringtonSynchronousDetectionCosmic2025,mcjunkinOnDemandCorrelatedErrors2026}, suggesting that a different mechanism mediates these transitions in JJ GE qubits.  Figure~\ref{fig:alphas}(a) displays the largest observed event in the JJ-Only dataset, determined by the matched filter trigger.  Figure~\ref{fig:alphas}(b) compares the correlated relaxation error rate integrated above various values for \Am and background datasets for both arrays. The background datasets were collected on the same arrays in the same cryostat during a different cooldown with no radiation source present.  Three of four datasets (the JJ-Only array with and without \Am, and the JJ\&M1 array without \Am) included 9 qubits.  The exception, the JJ\&M1 array dataset with \Am, includes 8 qubits --- two qubits in this array had nearly identical frequencies during this cooldown ($\Delta$\fq$<$\qty{1}{\mega\hertz}), preventing them from being independently addressed by the X$_\pi$ pulse.  To account for this, the matched filter output is normalized by the number of included qubits.  As a conservative lower bound for the observed $\alpha$-particle impact rate, we calculate the rate above the largest populated background bin for the JJ-Only array.  This rate is \qty[uncertainty-mode = separate]{0.064+-0.006}{\per\s} and matches the rate expected given the collimation geometry (\qty{0.077}{\per\s}) to within $2.2\sigma$.

Figure~\ref{fig:alphas}(c) shows the median qubit relaxation and excitation rates as a function of time, computed from our measurements using the procedure in Appendix~\ref{appendix:rates}.  As QPs in layer M2 thermalized to the superconducting gap cannot facilitate qubit transitions, we hypothesize that the QPs themselves have additional energy.  In this case, QPs in layer M2 can facilitate qubit relaxation so long as the expression

\begin{align}\label{eqn:relaxation_criteria}
E_\text{qp} > \delta\Delta_\text{JJ} - hf_\text{qb},
\end{align}

\noindent is satisfied, where $E_\text{qp}$ is the QP's energy relative to the superconducting gap.  Similarly, qubit excitation can occur for QPs in layer M2 if the following expression is satisfied 

\begin{align}\label{eqn:excitation_criteria_M2}
E_\text{qp} > \delta\Delta_\text{JJ} + f_\text{qb}.
\end{align}

\noindent QPs in layer M3 can facilitate both qubit relaxation and excitation so long as the expression

\begin{align}\label{eqn:excitation_criteria_M3}
\delta\Delta_\text{JJ}~>~hf_\text{qb},
\end{align}

\noindent is satisfied.

As support for this hypothesis, we approximate the QP energy distribution with an average qubit temperature ($T_\text{qb}$) during a particle interaction.  The temperature $T_\text{qb}$ is estimated to be~\cite{krantzQuantumEngineersGuide2019}:

\begin{equation}\label{eqn:temperature}
    T_\text{qb}\approx \frac{hf_\text{qb}}{k_B \mathrm{ln}(\frac{\Gamma_\downarrow}{\Gamma_\uparrow})}\quad,
\end{equation}

\noindent where $h$ is Plank's constant and $k_B$ is the Boltzmann constant.  The result of this estimation is shown in Fig.~\ref{fig:alphas}(d), where the qubit temperature is observed to transiently increase following the $\alpha$-particle impact and return to equilibrium over the course of approximately \qty{1}{ms}.  This highlights that although the relaxation rate largely recovers in hundreds of \qty{}{\us}, effects of radiation are present on the \qty{}{\ms} timescale.

\subsection{Implications for Future Processors}

While JJ GE significantly mitigates radiation-induced errors, qubit arrays protected in this way remain sensitive to radiation that deposits energy on the scale of a few~MeV.  
This indicates a sensitivity to energy depositions comparable to that of the hadronic component (protons and neutrons) of the cosmic ray spectrum.  We note that the 1-per-hour rate observed in Ref.~\onlinecite{acharyaQuantumErrorCorrection2025} is consistent with the rate of cosmic rays above approximately \qty{1}{\MeV}, where the radiation spectrum is primarily hadronic~\cite{fowlerSpectroscopicMeasurementsModels2024}.  Such hadronic cosmic rays can be mitigated by operation in shallow (approximately \qtyrange{15}{30}{\m}) underground facilities, where cosmic ray protons are attenuated with a mass attenuation thickness of approximately \qty{1.6}{\mwe} (meters of water equivalent) \cite{aalsethShallowUndergroundLaboratory2012,loerAbatementIonizingRadiation2024,bertoldoCosmicMuonFlux2025}.  Additionally, it highlights the importance of removing sources of $\alpha$-particle radiation (e.g., trace radioisotopes present in connectors and ceramics~\cite{loerAbatementIonizingRadiation2024, cardaniDisentanglingSourcesIonizing2023}) to avoid such correlated errors.  
This observed sensitivity to \qty{5.5}{\MeV} $\alpha$-particles motivates the development of further radiation mitigation strategies, discussed further in Section~\ref{sec:impact}. 


\section{\label{sec:linac} Qubit response to minimum-ionizing electrons}

\begin{figure*}[htbp]
    \begin{center}
        \includegraphics[width = \textwidth]{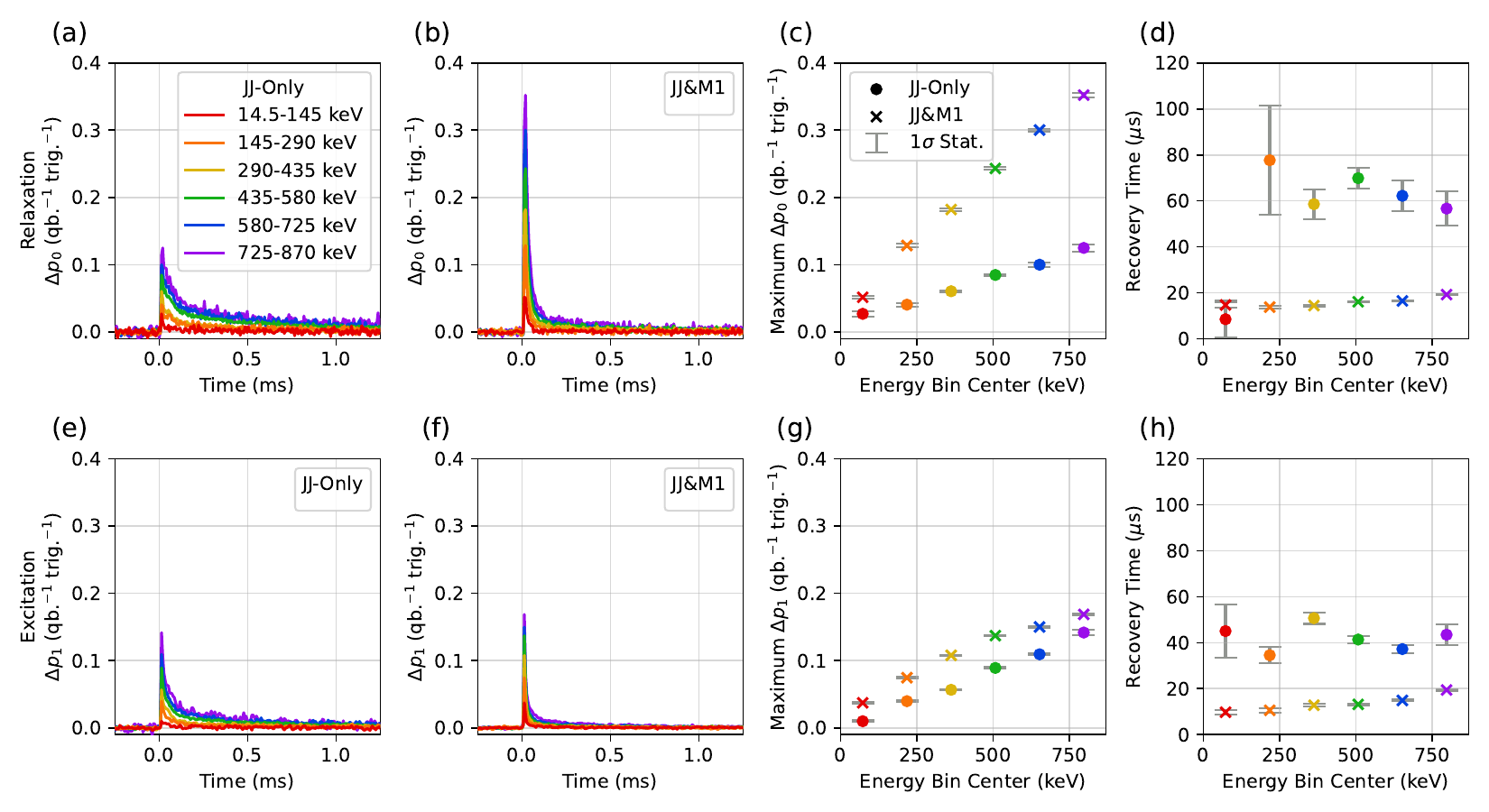}
    \end{center}
    \caption{Response of the qubit arrays to electrons at CLIQUE as a function of energy deposited in the qubit substrate.  The data are an average of approximately~5000 accelerator triggers taken in energy bins as labeled in panel (a). The top row (a-d) and bottom row (e-h) characterize qubit relaxation and excitation response respectively.  Panels (a, b, e, f) show the time-domain response to radiation events, with panels (a) and (e) showing data from the JJ-Only array, and panels (b) and (f) showing data from the JJ\&M1 array.  Panels (c) and (g) show the maximum measured change in qubit state probability, while panels (d) and (h) present the time constant of an exponential decay fit to the data in panels (a, b, e, f).  In panels (c, d, g, h) dots represent the JJ-Only array, crosses represent the JJ\&M1 array, and the $1\sigma$ statistical uncertainties are shown as gray vertical error bars.}
    \label{fig:linac_both_devices}
\end{figure*} 

\subsection{Measurements at CLIQUE}\label{linac_setup}

Precise measurements of qubit recovery dynamics and energy sensitivity were performed using CLIQUE~\cite{mcjunkinOnDemandCorrelatedErrors2026}.  This precision was facilitated by the accelerator's ability to time-tag $e^-$-qubit interactions and its trigger rate of \SI{10}{\hertz}, enabling us to rapidly collect and average together large samples of events.  We varied the mean number of electrons incident on the qubit substrate from \qtyrange{0.1}{6.9}{\electron}.  The accelerator current was monitored by an off-axis scintillator detector, correlated to electrons-on-target by a dedicated measurement where the qubit package was replaced with an electron detector~\cite{mcjunkinOnDemandCorrelatedErrors2026}.  The accelerated electrons have energies from \qtyrange{10}{20}{\mega\electronvolt} and therefore are minimum ionizing~\cite{navasReviewParticlePhysics2024}, allowing the average number of electrons-on-target to be converted to the energy deposited in the substrate.  The mean energy deposited was \SI[per-mode = symbol]{145}{\keV\per\electron} and the most probable energy deposited was \SI[per-mode = symbol]{96}{\keV\per\electron}.  The details of this conversion are presented in Ref.~\onlinecite{mcjunkinOnDemandCorrelatedErrors2026}.  Using this conversion, the range of measured accelerator currents corresponds to an average of \qtyrange[range-independent-prefix = true,exponent-to-prefix = true]{14.5e3}{1e6}{\electronvolt} deposited in the qubit substrate.

We performed five measurement sequences at CLIQUE~\cite{mcjunkinOnDemandCorrelatedErrors2026}.  To estimate qubit transition rates, we used sequences A, B, and D from Fig.~\ref{fig:sequences}.  Sequence D was used with two variants: with the conditional pulse set to prepare the qubit in \g (D$_{|0\rangle}$), and sequence D with the conditional pulse set to prepare the qubit in \e (D$_{|1\rangle}$).  To measure transient qubit frequency shifts, we used a repeated Ramsey interferometry sequence [Fig.~\ref{fig:detuning}(a)].  Further information regarding this interferometry sequence is presented in Sec.~\ref{sec:detuning}.

The qubit arrays were measured sequentially, each in a dedicated cryostat cooldown.  Within an array we measured up to nine qubits simultaneously.  The measurements were broken into basic units of ``runs.''  Each run comprised $10^7$ measurements and contained on the order of $10^3$ accelerator triggers, totaling approximately \SI{70}{\s} of data, with the specific value dependent on the measurement sequence used.  Appendix~\ref{appendix:linac} presents more details on the data collection process, with Fig.~\ref{fig:data_collected} showing the energy distribution of the data collected at CLIQUE. 

\subsection{Qubit Relaxation and Excitation}\label{sec:linac_T1}

We use measurement sequence D$_{|1\rangle}$ (D$_{|0\rangle}$) to estimate changes in the qubit relaxation (excitation) rate, estimated as the change in probability of measuring \g (\e). 

In Fig.~\ref{fig:linac_both_devices} we show the change in these state probabilities in response to electrons from CLIQUE as a function of the energy deposited in the qubit substrate.  The energy bins are described in panel (a), we use panels (a-d) to show qubit relaxation, and panels (e-h) to show qubit excitation.  Each energy bin is an average of approximately 5000 accelerator triggers.  Panels (a), (b), (e), and (f) show the average change in state probability as a function of time.  Panels (c) and (g) show the peak change in state probability.  Panels (d) and (h) show the recovery time, as estimated from a decaying exponential fit to data in panels (a), (b), (e), and (f).  We describe the details of the fit procedure in Appendix~\ref{appendix:linac}.  We hypothesize the exceptionally low recovery time reconstructed for the \qtyrange{14.5}{145}{\keV} JJ-Only array relaxation data is a fit to a noise fluctuation, due to the small signal size.

As with the \Am dataset, we observe a transient increase in qubit relaxation and excitation rate as a result of radiation depositing energy into the substrate.  As expected, the change in rate is larger for greater energies deposited in the qubit substrate [Fig.~\ref{fig:linac_both_devices}(c, g)].
Beyond these confirmations, a more striking observation is that qubits are affected by the lowest accelerator current measured: an average of \qtyrange{0.1}{1}{\electron} impinging on the Si, equivalent to an average of \qtyrange{14.5}{145}{\keV} deposited.  This indicates that the higher-rate (one per \qty{101}{\s} in Ref.~\onlinecite{harringtonSynchronousDetectionCosmic2025}), lower-energy radiation from cosmic ray muons and terrestrial $\gamma$-rays can still increase error probabilities in qubits strongly protected by JJ gap engineering.  
While radiation at these energy scales currently does not affect leading QEC experiments~\cite{mcewenResistingHighEnergyImpact2024,acharyaQuantumErrorCorrection2025}, a sensitivity will likely become apparent as qubit $T_1$ and $T_2$ improve.

Additionally, by comparing the JJ-Only and JJ\&M1 arrays, we observe that the degree of JJ GE (\ddjj/$h$\fq) matters.  Even though both arrays have \ddjj~$>h$\fq, the JJ-Only array has a $3\times$ lower peak change ground-state probability [Fig.~\ref{fig:linac_both_devices}(c)].  This is consistent with the qubit transitions being facilitated by hot QPs as suggested by the \Am dataset --- for the same QP energy distribution, more QPs satisfy the qubit relaxation condition of Eqn.~\ref{eqn:relaxation_criteria} and so the observed rate is greater.  A similar effect can be observed when measuring the excitation error rate [Fig.~\ref{fig:linac_both_devices}(g)].  The ratio of severity between the two gap profiles is different for excitation compared with relaxation.  We hypothesize this is because the energy required to facilitate qubit transitions differs between relaxation and excitation (cf., Eqns.~\ref{eqn:relaxation_criteria}~-~\ref{eqn:excitation_criteria_M3}).  A more quantitative discussion is presented in Section~\ref{sec:modeling}.

\begin{figure}[tbp]
    \begin{center}
    \includegraphics[width=0.7\columnwidth]{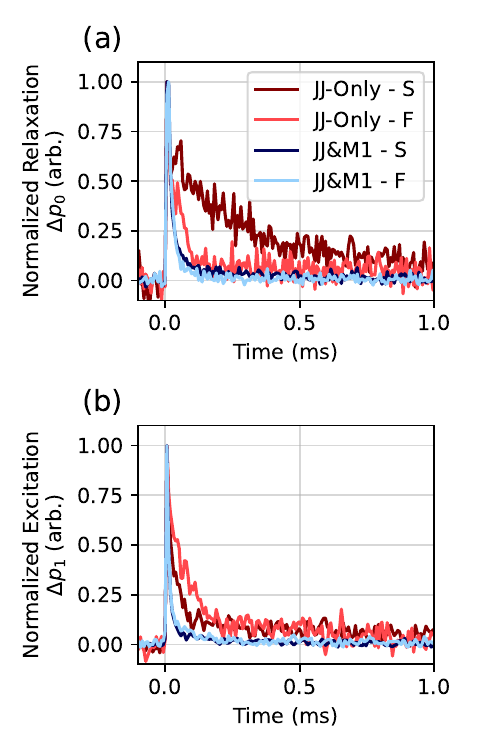}
    \end{center}
    \caption{Normalized change in state  in state probability of the qubit arrays in response to \qtyrange{145}{435}{\keV} energy depositions from electrons at CLIQUE.  Qubit response is organized by array (JJ-Only in warm colors, JJ\&M1 in cool colors), as well as by the two orientations (slow in dark colors, fast in light colors). Panel (a) shows qubit relaxation while (b) shows qubit excitation.  The data are normalized to a maximum value of one to highlight the differences in recovery timescales.}
    \label{fig:orientation_and_device}
\end{figure}

While the peak error rate is lower for the JJ-Only array, the radiation-induced transition rates remain elevated longer compared to the JJ\&M1 array [Fig.~\ref{fig:linac_both_devices}(d, h)].  We hypothesize that this is because the larger \ddgnd causes layer M1 to act as a QP trap, shortening the recovery time of the error rate.  This is supported by Fig.~\ref{fig:orientation_and_device} which normalizes the data to a peak value of one and separates the qubit response by JJ orientation.  Examining qubit relaxation for the JJ-Only array (low \ddgnd), we see a dramatic difference in response between the two orientations, an effect which can only be created by the influence of QPs in layer M1 (cf., QP dynamics solely in layers M2 and M3 should be agnostic to the JJ orientation).  The difference in response between orientations is significantly reduced for qubits in the JJ\&M1 array (high \ddgnd), indicating a diminished influence of QPs in layer M1 --- QPs are being trapped in layer M1.  
This implies that the introduction of a QP trap at the base metal interface (\ddgnd) is a complementary addition to JJ GE (\ddjj), and that both are important for suppressing qubit response to impacts from high-energy particles.

\subsection{Qubit Frequency Shifts}\label{sec:detuning}

\begin{figure}[tbp]
    \begin{center}
        \includegraphics[width=\columnwidth]{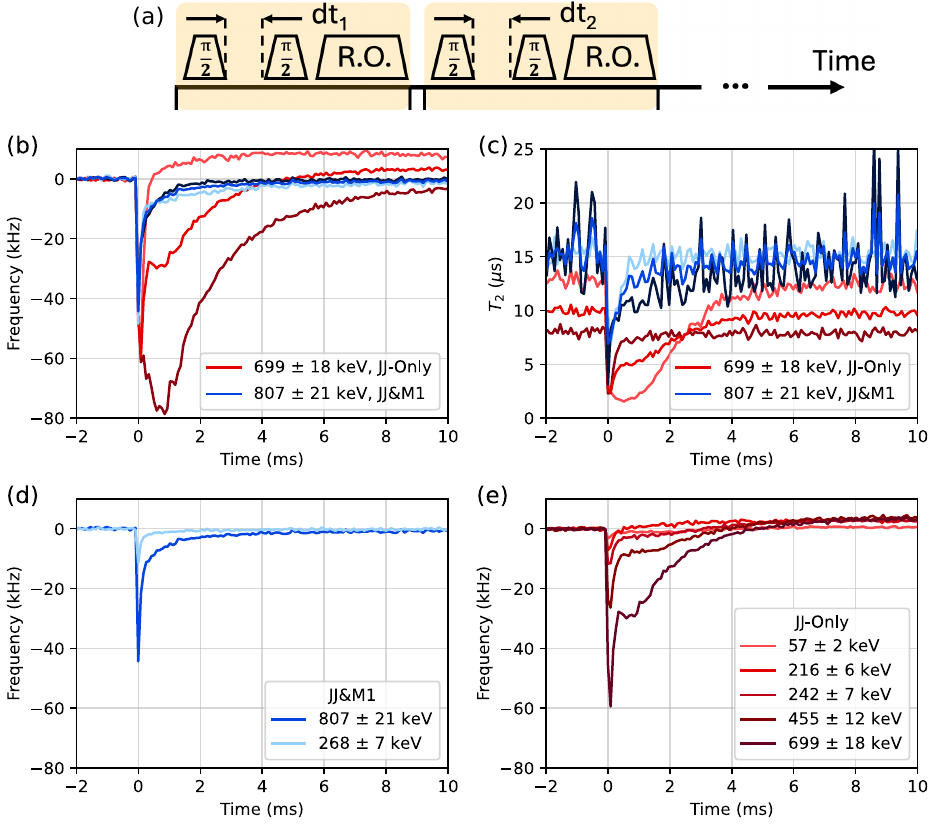}
    \end{center}
    \caption{Qubit frequency shifts in response to electrons from CLIQUE.  (a) The measurement sequence used for panels (b)-(e).  (b) Frequency shifts in response to the highest accelerator currents for each array, separated by JJ orientation.  The JJ-Only array is shown with warm colors and the JJ\&M1 array is shown with cool colors.  Dark and light colors represent slow and fast JJ orientations respectively, while the middle shade is the array average.  The JJ\&M1 array recovers more quickly, consistent with the fact that QPs are being trapped in the base metal by the large \ddgnd. (c) $T_2$ as a function of time since the accelerator trigger for each array.  Colors are identical to panel (b).  (d) The average frequency shift as a function of energy deposited in the substrate for the JJ\&M1 array, shades represent different energies.  (e) Average frequency shift as a function of energy deposited in the substrate for the JJ-Only array, shades represent different energies.}
    \label{fig:detuning}
\end{figure}

The final measurement sequence we employed at the accelerator targeted transient shifts in qubit frequency, which are expected to track the QP density near the JJ~\cite{catelaniRelaxationFrequencyShifts2011,kurilovichCorrelatedErrorBursts2025}.  This was accomplished by performing a Ramsey interferometry sequence with a set of 10 progressively longer delays, $\Delta t$, with step size of \qty{600}{\nano\s}~\cite{krantzQuantumEngineersGuide2019,mcjunkinOnDemandCorrelatedErrors2026} [Fig.~\ref{fig:detuning}(a)].  This set of measurements was repeated while the accelerator was fired, estimating the qubit frequency once for each 10-measurement series, equivalently once per approximately \qty{85}{\mu s}.  Resonant pulses were applied with a delay-dependent phase shift resulting in \qty{250}{\kHz} oscillations of the Ramsey signal, which was then fit to recover the frequency shift.  Details on the dataset construction, fit procedure, and data quality cuts are presented in Appendix~\ref{appendix:detuning}.

We show the resulting frequency shifts in Fig.~\ref{fig:detuning} and elaborate in Appendix~\ref{appendix:detuning}.  Consistent with previous work~\cite{kurilovichCorrelatedErrorBursts2025,mcjunkinOnDemandCorrelatedErrors2026} we see that the qubit frequency shifts in response to particle interactions~\cite{kurilovichCorrelatedErrorBursts2025}.  Unsurprisingly, we also observe that the frequency shift increases with larger energies deposited into the substrate.  

We note that the frequency shift recovers more quickly for qubits in the JJ\&M1 array than for those in the JJ-Only array.  Additionally, the frequency shifts of qubits in the JJ-Only array return to their equilibrium values nearly an order of magnitude more slowly than their relaxation rate [Fig.~\ref{fig:linac_both_devices}(d,h)].  As the qubit frequency shift is a proxy for the QP density near the JJ, this implies that QPs remain near the JJ in the JJ-Only array longer than the JJ\&M1 array.  Finally, we observe a transient reduction in qubit $T_2$ coinciding with the radiation-induced frequency shifts.

These observations lead to two implications.  First, QPs can remain near the JJ without causing relaxation or excitation errors.  This adds support to the hypothesis that the QP energy distribution is elevated following an radiation impact.  Second, the faster recovery time of qubits in the JJ\&M1 array implies that a gap difference at the ground-plane/JJ-lead interface \ddgnd traps QPs away from the JJ, as in Sec.~\ref{sec:linac_T1}.  This reinforces the importance of both \ddjj~and~\ddgnd for radiation resilience.  Further, we observe that for the fast JJ-orientation on the JJ-Only array the frequency shift increases relative to its nominal value, indicating that the QP density actually decreased for a portion of the transient event.  This behavior is a physically plausible result of a coupled system of Rothwarf-Taylor style differential equations [Ref.~\onlinecite{rothwarfMeasurementRecombinationLifetimes1967} or Sec.~\ref{sec:model_main_text}], with dramatically different effective volumes on either side of the JJ.  

\section{\label{sec:modeling} Modeling Quasiparticle Dynamics}

We now qualitatively compare the data collected at CLIQUE with a model of QP dynamics that captures key features in our data:

\begin{enumerate}
    \item The difference in severity and recovery timescale of qubit-state transition rates between the two gap engineering profiles, 
    \item The observed dependence of recovery timescales on JJ orientation, and
    \item The dependence of frequency shift measurements on gap profile and JJ orientation.
\end{enumerate}

\noindent The measured \g and \e state probabilities $p_0^m(t)$ and $p_1^m(t)$ are converted into excitation and relaxation transition rates $\Gamma_\uparrow(t)$ and $\Gamma_\downarrow(t)$ by combining our knowledge of the measurement sequences with a ``confusion matrix'' relating measured and true state probabilities.  This process is described in Appendix~\ref{sec:modeling_appendix}

\subsection{Model Construction}\label{sec:model_main_text}

Our model builds on the work in Refs.~\onlinecite{ diamondDistinguishingParitySwitchingMechanisms2022,connollyCoexistenceNonequilibriumDensity2024a,marchegianiNonequilibriumRegimesQuasiparticles2025,ql6q-wfpn}.  We begin by following \MG and consider the QP densities in the two metal layers adjacent to the JJ: M2 and M3.  To incorporate the gap profile near the JJ, the QP density in layer M2 is broken into two components: $x_{2<}$ for $E_{qp}<\delta\Delta_\text{JJ}$ and $x_{2>}$ for $E_{qp}>\delta\Delta_\text{JJ}$. We add two additional equations to the system presented in \MG to incorporate the effects of \ddgnd and the observed orientation dependence.  These additions $x_L$ and $x_R$ model the QP density in layers M1L and M1R (cf., Fig.~\ref{fig:devices}).  The resulting model is a system of six differential equations: one for the probability $p_0$ that the qubit occupies \g, and five others for the following QP densities

\begin{align}
    \Big\{x_i~\Big|~i \in \{2<,~2>,~3,~L,~R\}\Big\}.
\end{align}

\noindent As in \MG, the base structure of the equations for QP density is in the style of the ``Rothwarf-Taylor'' phenomenological model describing generation ($g$), linear trapping ($s x_i$), and quadratic recombination ($r x_i^2$)~\cite{rothwarfMeasurementRecombinationLifetimes1967}:

\begin{align}\label{eqn:RT}
    \dot{x}_i = g - s x_i - r x_i^2.
\end{align}

\noindent The QP densities in layer M1 serve as source terms for the QP densities in layers M2 and M3.  We assume the QP density in layer M1 behaves independently of the QP population in M2 and M3 due to the volume difference --- a QP transitioning between M1 and either M2 or M3 has little effect on the QP densities in the M1. Additionally, the QP densities in layers M2 and M3 are coupled to each other and to qubit transitions.  This coupling to qubit transitions allows us to compare the model to the observed excitation and relaxation rates on a per-qubit basis.  The full system of equations is shown in Appendix~\ref{sec:modeling_appendix} along with a more detailed discussion of the model construction.  

The transient QP energy distribution is parametrized by a time-dependent, spatially homogeneous temperature $T(t)$ describing the phonon bath.  The time dependence is modeled with one exponential rise-time and two exponential fall-times.  The functional form is presented in Eqns.~\ref{eqn:temp}~-~\ref{eqn:deltaT}.  As mentioned in Sec.~\ref{sec:am_source}, this transient increase in effective temperature provides intuition for the physical mechanism mediating these state transitions: the QPs themselves have elevated energy.  Energetic QPs can cause qubit relaxation or excitation in Eqns.~\ref{eqn:relaxation_criteria}~-~\ref{eqn:excitation_criteria_M3} are satisfied~\cite{serniakHotNonequilibriumQuasiparticles2018}.

\subsection{Comparing the Model and Data}\label{sec:model_comparison}

\begin{figure}[tbp]
    \begin{center}
    \includegraphics[width=\columnwidth]{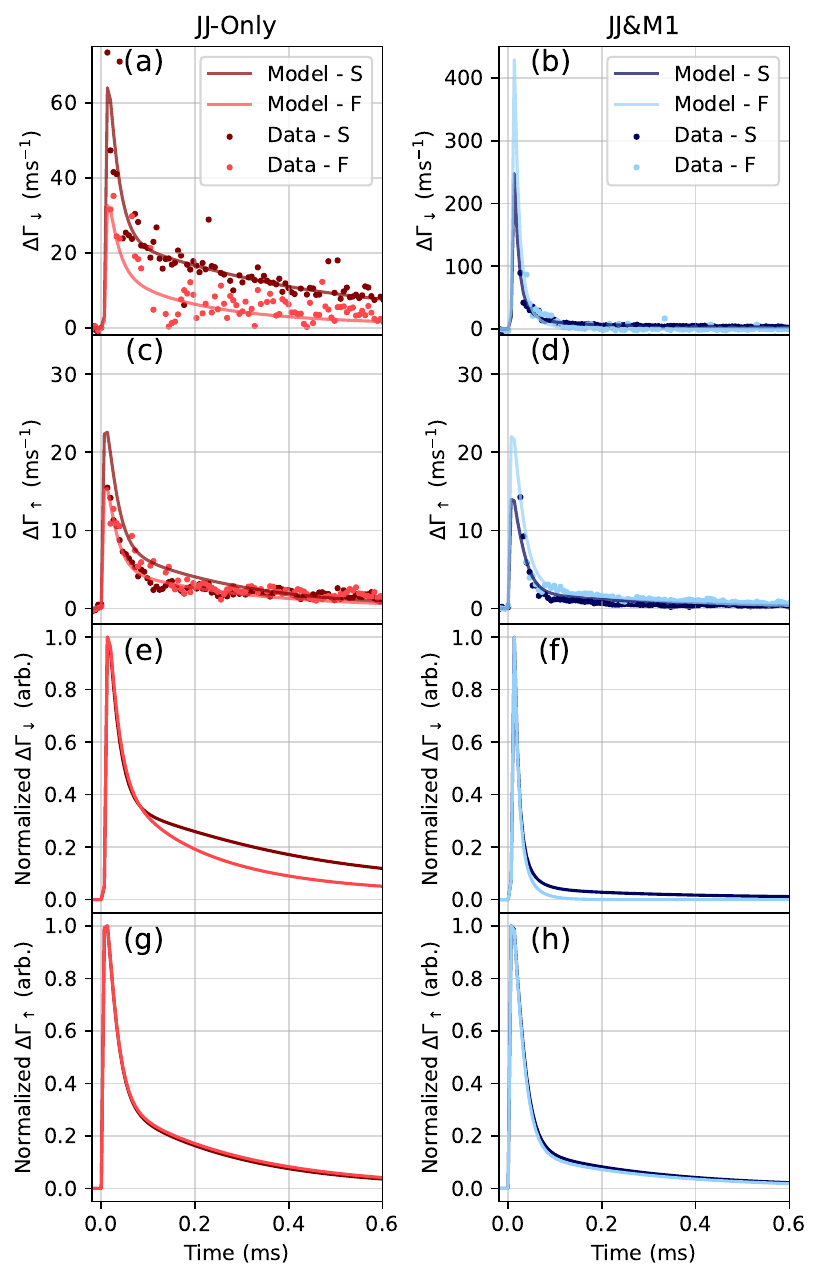}
    \end{center}
    \caption{Average transition rates (dots) and comparison with the QP model (lines).  We show the average change in relaxation and excitation rate for the JJ-Only array in panels (a), (c), (e), and (g).  Dark red and light red represent slow and fast JJ orientations respectively.    We show the average change in relaxation and excitation rate for the JJ-Only array in panels (b), (d), (f), and (h).  Dark blue and light blue represent represent slow and fast JJ orientations respectively.  Neutral blue shows the array average.  Panels (e)-(h) show the model from (a)-(d), except normalized to a peak value of one for comparison with Fig.~\ref{fig:orientation_and_device}.  Panels (a), (b), (e), and (f) show the qubit relaxation rate, while panels (c), (d), (g), and (h) show the qubit excitation rate.}
    \label{fig:transition_rates}
\end{figure}

Figure~\ref{fig:transition_rates} presents $\Gamma_\uparrow$ and $\Gamma_\downarrow$ averaged over each array and delineated by JJ orientation.  Data are shown in dots, while the model is overlaid as solid lines using physically motivated values for the parameters --- the model is not fit to the data due to the large model parameter space.  A discussion of the parameters used and further details of this overlay are contained in Appendix~\ref{sec:modeling_appendix}.  A breakdown of Fig.~\ref{fig:transition_rates} into individual qubits is shown in Fig.~\ref{fig:modeling_by_qubit} of Appendix~\ref{appendix:model_and_data}.  The qubit on resonator 4 of the JJ-Only array is excluded from averages due to its extraordinarily large baseline relaxation rate (greater than \qty{125}{\per\milli\s}).  

We observe general consistency between the model and data --- we are able to simultaneously reconstruct the overall magnitude and recovery timescale for both relaxation and excitation in both arrays (a total of 17 qubits).  As in the discussion of Sec.~\ref{sec:linac}, the modeled JJ\&M1 array has a larger peak relaxation rate but faster recovery timescale compared to JJ-Only.  Additionally, the model qualitatively recreates the behavior for both the slow and fast JJ orientations.  This is highlighted in Fig.~\ref{fig:transition_rates}(e)-(h), where the response has been normalized akin to Fig.~\ref{fig:orientation_and_device}.  We see the second, slower fall-time for qubit relaxation is reduced for qubits with a fast JJ orientation, and as in data the effect is larger for qubits in the JJ-Only array [Fig.~\ref{fig:transition_rates}(e)-(f)].  Additionally, the orientation dependence for qubit excitation is small [Fig.~\ref{fig:transition_rates}(g)-(h)].

\begin{figure}[tbp]
    \begin{center}
    \includegraphics[width=0.8\columnwidth]{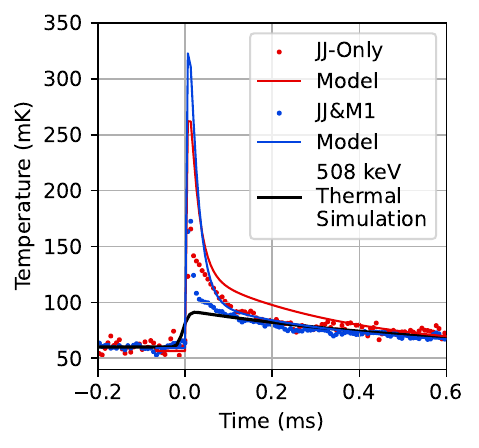}
    \end{center}
    \caption{Average qubit temperature in response to a \qty{508}{\keV} energy deposition.  The dots represent data, calculated from the qubit transition rates in Fig.~\ref{fig:transition_rates} with Eqn.~\ref{eqn:temperature}.  This is compared with the temperature profiles used in the modeling (lines) and a non-linear thermal simulation (black).  The JJ-Only array is shown in red and the JJ\&M1 array is shown in blue.}
    \label{fig:temperature}
\end{figure}

Figure~\ref{fig:temperature} shows the qubit temperature evolution for the JJ-Only (red) and JJ\&M1 (blue) arrays, including both the temperature computed with Eqn.~\ref{eqn:temperature} and the temperature profile used in the modeling.  The data peak at approximately \qty{175}{\milli\K} and exhibit two exponential fall times, motivating the functional form used for $T(t)$.  In addition, we perform a non-linear thermal simulation~\cite{figueroa-felicianoComplexMicrocalorimeterModels2006, bastidonOptimizingThermalDetectors2018, chenTransitionEdgeSensor2022, chenModelingCharacterizationTESBased2024} of energy depositions in our qubit substrate, thermalized by the aluminum wire bonds from the ground plane base metal to the PCB.  This thermal simulation is identical to that shown in Fig.~\ref{fig:alphas}(d), except for a \qty{508}{\keV} (\qty{3.5}{\electron}) instead of \qty{5.5}{\MeV}.  With no free parameters and assuming both a starting temperature of \qty{60}{\milli\K} and an average wire bond length of \qty{500}{\um}, this model (black line) reconstructs the magnitude and timescale of the slower fall-time present in our data.  We hypothesize that the fast timescale before \qty{100}{\mu s} is due to the effect of athermal phonons and do not expect to reconstruct this feature with our simulation.

\begin{figure}[tbp]
    \begin{center}
        \includegraphics[width=0.8\columnwidth]{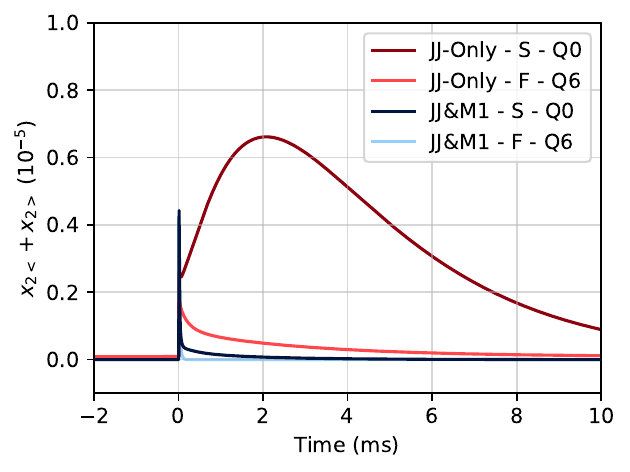}
    \end{center}
    \caption{Modeled reduced QP density in layer M2, the sum of $x_{2<}$ and $x_{2>}$, for representative qubits in each array.  The model predicts qualitatively similar timescale relative to the frequency shits observed in Fig.~\ref{fig:detuning}.  A representative example of each qubit gap profile and JJ orientation is presented, with reds representing JJ-Only and blues representing the JJ\&M1 arrays.  The slow JJ orientation is shown in dark colors while fast JJ orientation is in light colors.}
    \label{fig:qp_densities}
\end{figure}

Finally, for comparison with the frequency shift measurements in Fig.~\ref{fig:detuning}, we display the reduced QP density in layer M2 as predicted by our model in Fig.~\ref{fig:qp_densities}.  With the exception of the ``overshoot'' observed in the fast orientation qubits in the JJ-Only array, our model qualitatively re-creates the observed behavior, predicting a millisecond-scale recovery time for QPs in layer M2 for slow JJ orientation qubits on the JJ-Only array and sub-millisecond recovery time for the remaining qubits.  As mentioned in Sec.~\ref{sec:linac}, the ``overshoot'' behavior is a plausible solution to a coupled system of Rothwarf-Taylor differential equations, though it is not predicted by the specific parameters used in this model.

Multiple approximations have been made in this analysis.  In terms of the model, physical processes such as QP diffusion are not included.  Additionally, the gap engineering parameters used are the nominal, designed values.  While these values were extracted from transition temperature measurements on films from the same fabrication process, they were not measured on these individual chips.  In terms of data, averaging over accelerator triggers also averages over impacts across the entire qubit substrate.  We expect qubits further from the interaction point to exhibit a weaker response~\cite{yeltonModelingPhononmediatedQuasiparticle2024,mcewenResolvingCatastrophicError2022}, and so averaging could smear out this effect and distort our measured qubit responses.  Finally, despite our efforts to account for qubit states $\geq|2\rangle$ through the procedure described in Appendix~\ref{appendix:linac_rates}, the transformation between measured probabilities and transition rates fundamentally assumes a two-level system, which could additionally distort data.  Future improvements to this modeling work could be to combine it with radiation and condensed matter simulation packages such as GEANT4~\cite{allisonRecentDevelopmentsGeant42016} and G4CMP~\cite{kelseyG4CMPCondensedMatter2023,yeltonModelingPhononmediatedQuasiparticle2024,PhysRevD.111.063047,larsonQuasiparticlePoisoningSuperconducting2025}.


\section{\label{sec:impact} Impact and Conclusions}

This work reinforces the assertion that JJ GE superconducting qubits are still negatively affected by radiation.  
In particular, our measurements of MeV-scale energy depositions from $\alpha$-particles support a hypothesis that hadronic cosmic rays are a major contributor to the \qty{1e-10}{} error floor observed in Ref.~\onlinecite{acharyaQuantumErrorCorrection2025}.  This is further substantiated by the similarity of the recovery timescale between our work and Ref.~\onlinecite{acharyaQuantumErrorCorrection2025}, and the fact that the correlated error rate observed in Ref.~\onlinecite{acharyaQuantumErrorCorrection2025} is comparable to the expected impact rate of cosmic rays above \qty{1}{\MeV}~\cite{fowlerSpectroscopicMeasurementsModels2024}.
Unlike cosmic ray muons, these hadronic cosmic rays can be mitigated by operation in a shallow (\qtyrange{15}{30}{m}) underground facility~\cite{aalsethShallowUndergroundLaboratory2012,loerAbatementIonizingRadiation2024,bertoldoCosmicMuonFlux2025}.  While JJ GE is a significant step towards mitigating correlated errors from radiation, additional error mitigation strategies will be required in quantum processors depending on the desired performance.  

In addition, $\alpha$-particles could be expected to become significant in future quantum computers.  These high-energy particles can be created by, for example, lightly radioactive components~\cite{loerAbatementIonizingRadiation2024} or, as observed in rare-event-search particle physics experiments, gaseous radon on surfaces~\cite{akeribLUXZEPLINLZRadioactivity2020,supercdmscollaborationStrategyLowMassDark2022,thelux-zeplincollaborationBackgroundDeterminationLUXZEPLIN2023}.  Should $\alpha$-particle backgrounds become significant, they can be mitigated by ensuring that materials with line-of-sight to the qubit substrate are radio-pure and radon-free --- $\alpha$-particles have a range of tens of microns in most materials: (approximately \qty{28}{\mu m} in silicon and \qty{11}{\mu m} in copper at \qty{5.5}{\MeV}~\cite{148751}).

Beyond these high-energy-particle-induced correlated errors, our measurements at CLIQUE show that sub-MeV radiation, such as terrestrial $\gamma$-rays and cosmic ray muons, can still cause state transitions in gap-engineered superconducting qubit arrays.  The effect of these particles could become apparent as non-radiation sources of error are reduced.  While the high rate of these impacts would pose a challenge, the spatial extent of their effect on gap-engineered qubits (and therefore the extent of their effect on QEC) remains unknown.  The accelerator data presented here cannot answer these questions as accelerator electron impacts are distributed evenly across the qubit arrays under study, and the data are heavily averaged.  These correlations can be studied in detail with a combination of co-located sensors~\cite{orrellSensorAssistedFaultMitigation2021} (e.g., microwave kinetic inductance detectors~\cite{dayBroadbandSuperconductingDetector2003}, transition-edge sensors~\cite{irwinQuasiparticletrapassistedTransitionedgeSensor1995}, or qubits with minimal radiation protection or enhanced radiation sensitivity~\cite{magoonFirstDemonstrationSQUAT2026,sundelinRealtimeDetectionCorrelated2026}) and low-energy radiation sources with improved collimation and tagging.

Further, this work highlights the importance of understanding radiation's impact on qubits beyond $T_1$.  As previously observed~\cite{kurilovichCorrelatedErrorBursts2025,mcjunkinOnDemandCorrelatedErrors2026}, we identify significant qubit-frequency shifts in response to radiation impacts, although they are reduced relative to previous measurements~\cite{kurilovichCorrelatedErrorBursts2025}.  The reduction with respect to Ref.~\onlinecite{kurilovichCorrelatedErrorBursts2025} is likely due to material and gap-engineering differences between measurements.  We also identified a transient decrease in qubit $T_2$ following a radiation impact.

Additionally, we identify that both the degree of JJ GE (\ddjj/$h$\fq) and the presence of a gap difference at the M1 layer interface (\ddgnd) influence qubit recovery from radiation impacts.  While larger \ddjj suppresses the initial elevation in qubit transition rate, larger \ddgnd hastens qubit recovery.  The effect of \ddgnd is most noticeable on the recovery of qubit frequency, where the increased \ddgnd reduced the recovery times by an order of magnitude.  With the knowledge that both \ddjj and \ddgnd influence radiation resistance, future work could focus on maximizing both gap differences and employing other QP-trap and phonon-sink designs~\cite{yeltonModelingPhononmediatedQuasiparticle2024,larsonQuasiparticlePoisoningSuperconducting2025}. In particular, we highlight a disadvantage of combining high-gap base metal with aluminum-based JJs~\cite{altoeLocalizationMitigationLoss2022,blandMillisecondLifetimesCoherence2025} --- the negative \ddgnd turns the JJ into a QP trap, amplifying the impact of radiation and QPs as a whole.  As a consequence, this work motivates the pursuit of higher gap JJs (e.g., NbN/AlN and Nb/Al/AlO$_\text{x}$)~\cite{gurvitchHighQualityRefractory1983,kimEnhancedCoherenceAllnitride2021,anferovImprovedCoherenceOptically2024} and improvements to the film quality of lower-gap base metal materials (e.g., Al, Re, and material stacks utilizing the proximity effect~\cite{serniakNonequilibriumQuasiparticlesSuperconducting2019}).

Finally, we developed and implemented a time-dynamic model of QP density evolution following energy deposition by high-energy radiation.  We observed qualitative agreement between this model and our data.  This model could serve as the basis for future investigations into how QP and phonon dynamics affect superconducting circuits, and could potentially be incorporated into packages such as G4CMP~\cite{kelseyG4CMPCondensedMatter2023}. 

\begin{acknowledgments}

This research is sponsored by the U.S. Army Research Office under Award No. W911NF-23-1-0045 (Extensible and Modular Advanced Qubits), and under Air Force Contract No. FA8702-15-D-0001. The views and conclusions contained herein are those of the authors and should not be interpreted as necessarily representing the official policies or endorsements, either expressed or implied, of the U.S. Government.  The authors acknowledge JHU/APL for the infrastructure investments required to perform these experiments, and conversations with Joseph Fowler regarding the natural radiation spectrum affecting qubits.

\end{acknowledgments}

\appendix

\include{appendix}

\bibliography{apl_measurement_bib_20260116}

\end{document}

%% file: alpha_linac_authorlist.tex
\newcommand{\rle}{\affiliation{Research Laboratory of Electronics, Massachusetts Institute of Technology, Cambridge, Massachusetts 02139, USA}}
\newcommand{\lns}{\affiliation{Laboratory for Nuclear Science, Massachusetts Institute of Technology, Cambridge, Massachusetts 02139, USA}}
\newcommand{\lincoln}{\affiliation{Lincoln Laboratory, Massachusetts Institute of Technology, Lexington, Massachusetts 02421, USA}}
\newcommand{\jhuapl}{\affiliation{Johns Hopkins Applied Physics Laboratory, Laurel, Maryland 20723, USA}}
\newcommand{\phys}{\affiliation{Department of Physics, Massachusetts Institute of Technology, Cambridge, Massachusetts 02139, USA}}
\newcommand{\eecs}{\affiliation{Department of Electrical Engineering and Computer Science, Massachusetts Institute of Technology, Cambridge, Massachusetts 02139, USA}}
\newcommand{\mines}{\affiliation{Department of Physics, Colorado School of Mines, Golden, Colorado 80401, USA}}

\author{H.~D.~Pinckney} \email{hdpinck@mit.edu}  \rle\lns
\author{T.~McJunkin} \jhuapl
\author{A.~W.~Hunt} \jhuapl
\author{P.~M.~Harrington} \rle\lns
\author{H.~P.~Binney} \rle\lns
\author{M.~Hays} \rle
\author{Y.~Jones-Alberty} \jhuapl
\author{K.~Azar} \rle
\author{F.~Contipelli} \lincoln
\author{R.~DePencier Piñero} \lincoln
\author{J.~M.~Gertler} \lincoln
\author{M.~Gingras} \lincoln
\author{A.~Goswami} \rle
\author{C.~F.~Hirjibehedin}\lincoln
\author{M.~Li}\lns
\author{M.~Moes}\rle
\author{B.~M.~Niedzielski} \lincoln
\author{M.~T.~Randeria} \lincoln
\author{R.~Sitler} \jhuapl
\author{M.~K.~Spear} \jhuapl
\author{H.~Stickler} \lincoln
\author{J.~Yang} \lns
\author{W.~Van~De~Pontseele} \mines 
\author{M.~E.~Schwartz} \lincoln
\author{J.~A.~Grover} \rle
\author{K.~Schultz} \jhuapl
\author{K.~Serniak} \rle\lincoln
\author{J.~A.~Formaggio} \lns\phys
\author{W.~D.~Oliver} \rle\phys\eecs

%% file: appendix.tex
\begin{widetext}

\section{Data Collection}

\subsection{Measurement Sequences}\label{sec:appendix_measurement_sequences}

Changes in qubit excitation and relaxation rates were identified with the four measurement sequences shown in Fig.~\ref{fig:sequences}.  

Measurements of \Am utilized sequences A-C.  Sequence A always applies a X$_\pi$ pulse between qubit measurements.  This increases the \e state population, and enables a measurement primarily sensitive to qubit relaxation.  Sequence B never applies a X$_\pi$ pulse between measurements.  This emphasizes the \g state population, enabling a measurement primarily sensitive to qubit excitation.  Sequence C alternates between A and B on a shot-by-shot basis.  This enables near simultaneous measurement of qubit excitation and relaxation.  The time between successive qubit measurements in sequences A, B, and C was \qty{6.95}{\micro\second} for measurements with \Am.  We refer to this time as the ``cadence'' of the sequence.

For measurements carried out at CLIQUE, we added a fourth measurement sequence, D, in addition to sequences A and B.  Sequence D performs an ``active reset'', using the previous measurement to decide whether or not to apply the X$_\pi$ pulse.  This sequence was run with the conditional logic targeting either \g or \e (D$_{|0\rangle}$ and D$_{|1\rangle}$ respectively).  At CLIQUE, sequences A and B had a cadence of \qty{6.55}{\micro\second}, and sequence D had a cadence of \qty{7.43}{\micro\second}.  We did not use sequence C for measurements at CLIQUE.

\subsection{Setup for Measurements with \Am}\label{appendix:am-241}

Figure~\ref{fig:am-241} provides an illustration of the measurement setup for the \Am measurement. When collecting data for the JJ-Only array, the source was collimated using an approximately \SI{100}{\micro\m} diameter pinhole in copper tape, and when collecting data for the JJ\&M1 array, the source was collimated using an \qty{7.96}{\milli\meter} thick custom machined block of Cu101 with three \qty{180}{\um} diameter holes.

\begin{figure}[tbp]
    \begin{center}
        \includegraphics[width=0.35\columnwidth]{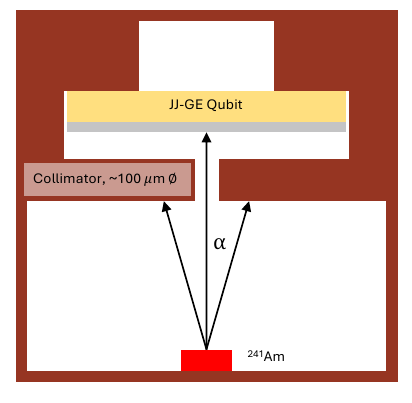}
    \end{center}
    \caption{Illustration of the measurement setup for measurements with \Am.  Not to scale.  The copper housing (brown) contains a qubit chip with a silicon substrate (yellow) and aluminum metal (gray).  Alpha particles (black lines) from the \Am source (red) are collimated before impinging on the qubit array.}
    \label{fig:am-241}
\end{figure}

\subsection{CLIQUE Dataset Details}\label{appendix:linac}

The frequency and $T_1$ of qubits is summarized in Table~\ref{tab:t1_and_freq} for data collected at CLIQUE.  Resonators are indexed by their frequency.

Figure~\ref{fig:data_collected} presents the distribution of data collected at CLIQUE for each array as a function of the average number of electrons incident on the qubit substrate.  This figure only presents data collected using sequences A, B, D$_{|0\rangle}$, and D$_{|1\rangle}$.  Data was evenly distributed between these four sequences.

We cycled through the four sequences when taking the transition rate datasets.  We recalibrated the qubit frequencies and X$_\pi$ pulse amplitudes every eight runs (two cycles of the four measurements).  Following each recalibration, we shifted the measurement order such that a different sequence was measured first; this way, all sequences were measured with an even distribution of time-since-recalibration.  Occasionally, a qubit's calibration parameters would vary significantly between and within entries.  Such qubits were temporarily excluded from the data collection.  

The fits presented in Fig~\ref{fig:linac_both_devices}(d,h) were performed on data from \qtyrange{19}{203}{\us} following an accelerator trigger (\qtyrange{13}{190}{\us} following the typical time of maximum probability change).  This delay ensures the longer decay timescale is captured by the fit.   Data was fit to a decaying exponential with a constant offset, and least-squares minimization was done in \texttt{Python} with \texttt{iminuit}'s \texttt{migrad} algorithm.

Measurements targeting qubit frequency shifts are discussed in the main text in Section~\ref{sec:detuning}.

\begin{figure}[tbp]
    \begin{center}
    \includegraphics[width=0.4\columnwidth]{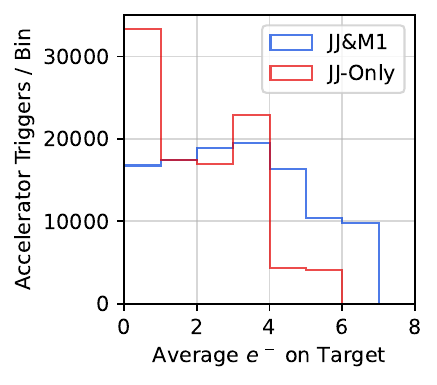}
    \end{center}
    \caption{Number of accelerator triggers (electron-qubit interactions) sampled as a function of energy for transition rate measurements at CLIQUE.  Includes sequences A, B, D$_{|0\rangle}$, and D$_{|1\rangle}$}
    \label{fig:data_collected}
\end{figure}

\begin{table}[tbp]
\centering
\begin{tabular}{ccccc}
\toprule
Resonator & \multicolumn{2}{c}{JJ-Only} & \multicolumn{2}{c}{JJ\&M1}\\
 Index & $T_1$ ($\mu$s) & \fq (GHz) & $T_1$ ($\mu$s) & \fq (GHz) \\
\midrule
0 & $45.1 \pm 6.5$ & 4.928 & $44.0 \pm 7.2$ & 4.704 \\
1 & $40.1 \pm 10.4$ & 4.919 & $55.1 \pm 9.0$ & 4.912 \\
2 & $34.4 \pm 5.2$ & 5.035 & $30.2 \pm 4.4$ & 5.164 \\
3 & $37.5 \pm 6.6$ & 5.031 & $39.2 \pm 8.0$ & 5.009 \\
4 & $5.7 \pm 0.8$ & 5.202 & --- & --- \\
5 & --- & --- & $29.9 \pm 6.0$ & 5.211 \\
6 & $28.9 \pm 8.6$ & 5.380 & $22.3 \pm 3.3$ & 5.502 \\
7 & $29.7 \pm 4.0$ & 5.480 & $20.0 \pm 2.1$ & 5.433 \\
8 & $25.2 \pm 3.9$ & 5.563 & $22.0 \pm 3.4$ & 5.461 \\
9 & $21.9 \pm 2.7$ & 5.734 & $24.6 \pm 2.9$ & 5.468 \\
\bottomrule
\end{tabular}
\caption{Median energy relaxation timescale $T_1$ and qubit frequency \fq for each qubit measured qubit at CLIQUE.  Uncertainty on $T_1$ represents the standard deviation of the measured distribution, collected over multiple days.  The $T_1$ times during measurements with \Am are consistent with these values.  We dropped two qubits (one on each array) as only nine qubits could be measured simultaneously.}\label{tab:t1_and_freq}
\end{table}

\section{Alpha Particle Rate Analysis Data Selection}\label{appendix:alpha_data_selection}

The time series shown in Fig.~\ref{fig:alphas}(c,d) are an average over multiple radiation impacts identified through the following process.  Following data collection, relaxation and excitation errors are computed following the logic described in Sec.~\ref{sec:am_source}.  The response of all qubits is then summed, and then divided by the number of qubits where first measurement was \g (removing qubits with ambiguous state preparation).  This yields a time series dataset of ``number of qubit transitions per measurement per valid qubit'' --- an error probability.  This is filtered by cross-correlating with a \qty{100}{\us} decaying exponential, a ``matched filter''~\cite{harringtonSynchronousDetectionCosmic2025}, to increase the signal-to-noise ratio of radiation impacts.  

To further improve our sensitivity to radiation impacts we perform a simple event-shape estimation by comparing the peak value found in the matched filter with the integral of the first 15 data points following the matched filter peak.  The expectation is that radiation should have an integral that scales more strongly with matched filter height than noise fluctuations.  This is what we observe [Fig.~\ref{fig:alpha_data_selection}].  To conservatively select radiation impacts, we remove all events where the integral of the qubit relaxation signal is less than 7.4.

\begin{figure}[tbp]
    \begin{center}
        \includegraphics[width=0.5\columnwidth]{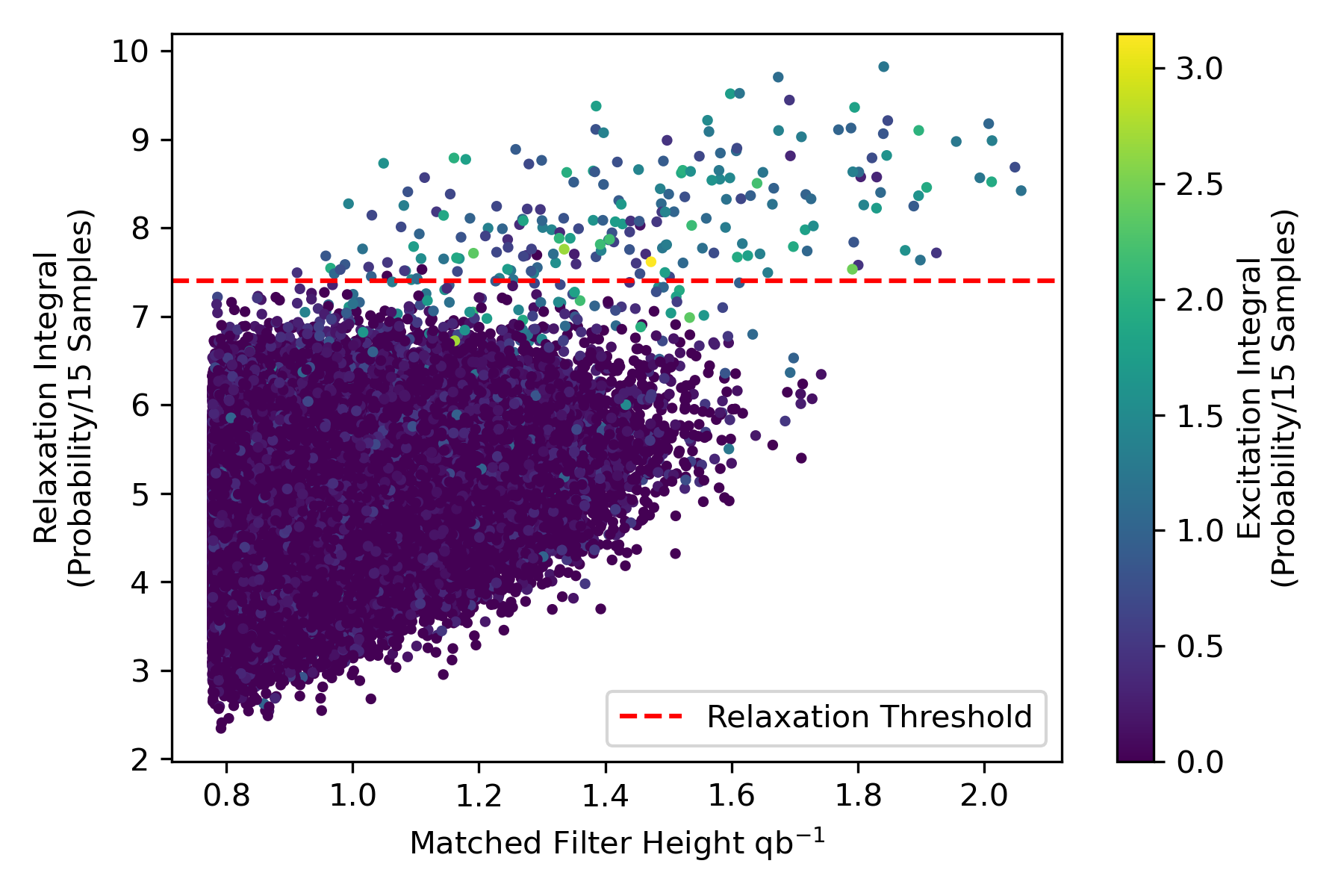}
    \end{center}
    \caption{Data selection for Fig.~\ref{fig:alphas}(c,d).  Event shape is estimated by comparing matched filter score and transition probability integral.  Events with an integral below a relaxation probability integral of 7.4 (dashed red line) are cut from the analysis.}
    \label{fig:alpha_data_selection}
\end{figure}

\section{Qubit Frequency Shifts}\label{appendix:detuning}

We estimate the qubit frequency through the 10-point Ramsey sequence described in Sec.~\ref{sec:detuning}.  The data are fit to the following expression:

\begin{equation}\label{eqn:detuning}
    p_0(\Delta t) = ae^{-t/\tau}\mathrm{sin}(2\pi f_\text{qb}\Delta t + \phi) + b,
\end{equation}

\noindent where $p_0$ is the probability of \g, $a$ and $b$ are constants, $\tau$ is the decay time constant, and $\phi$ is the phase.  The dataset used to fit Eqn.~\ref{eqn:detuning} is constructed as follows.  The measurement closest to the accelerator trigger is identified and the data near in time to the trigger is broken into sets of 10 measurements (one for each Ramsey sequence $\Delta t$).  The 10 measurements are arranged according to $\Delta t$ and then averaged over many accelerator triggers.  While fitting, in the time before the accelerator trigger ($t<0$ in Fig.~\ref{fig:detuning}) the parameters $a$, $\tau$, \fq, $\phi$, and $b$ are allowed to vary.  Immediately following the accelerator trigger ($t>0$ in Fig.~\ref{fig:detuning}) we allow $a$, $\tau$, \fq, and $b$ to vary while $\phi$ is fixed to the steady-state value.  

In averages, qubits with a reconstructed $T_2$ less than \qty{5.4}{\micro\s} are removed from the analysis, as this is shorter than the longest measurement delay.  Additionally, the qubit on resonator 3 of the JJ-Only array is removed as the software detuning was much greater than intended, leading to an inability to properly reconstruct the qubit frequency.

Figure~\ref{fig:detuning_supplement} provides additional information for Fig.~\ref{fig:detuning} in the main text.  Figure~\ref{fig:detuning_supplement}(a) shows representative fits at four selected times at 20-sample intervals for the dataset with the highest average energy deposited: quiescent value pre-trigger, interaction peak, 20 samples after trigger, and 40 samples after trigger.  It is clear from these examples that a shift in both qubit frequency and $T_2$ is observed.

\begin{figure}[tbp]
    \begin{center}
        \includegraphics[width=0.35\columnwidth]{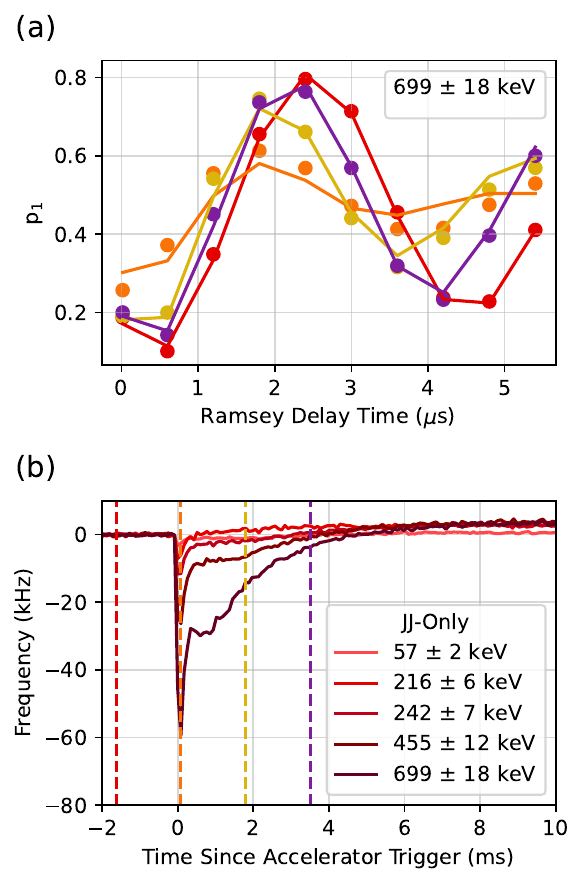}
    \end{center}
    \caption{(a) Example data (points) and fits (lines) for four time points on the JJ-Only device.  Time points are indicated on the bottom right with dashed vertical lines of corresponding color.  (b) Average frequency shift as a function of energy deposited in the substrate for the JJ-Only device.  Data identical to Fig.~\ref{fig:detuning}(d).  Lines indicate the times of the data points shown in panel (a).}
    \label{fig:detuning_supplement}
\end{figure}

\section{Calculation of Transition Rates}\label{appendix:rates}

To calculate transition rates and confusion matrix parameters from our data, we begin by writing the time evolution of qubit state probabilities from time $t$ to $t + \Delta t$:

\begin{equation}
    \vec{p}^{~a}(t + \Delta t) = \hat{D}(\Delta t,\Gamma_\uparrow,\Gamma_\downarrow)~ \vec{p}^{~a}(t),
\end{equation}

\begin{equation}
    \vec{p}^{~a}(t) = \begin{pmatrix}
                    p_{0}^a(t)\\
                    p_{1}^a(t)
            \end{pmatrix} ,
\end{equation}

\noindent where $p_{0}^a$ and $p_{1}^a$ are the true ground and excited state probabilities, and the time evolution is given by the transition matrix $\hat{D}(\Delta t,\Gamma_\uparrow,\Gamma_\downarrow)$, defined as:

\begin{align}\label{eqn:transition_matrix}
\hat{D}(\Delta t) =
\begin{pmatrix}
T_1 (\Gamma_\uparrow e^{-\Delta t/T_1} + \Gamma_\downarrow ) & 1-T_1(\Gamma_\downarrow e^{-\Delta t/T_1} + \Gamma_\uparrow) \\
1-T_1 (\Gamma_\uparrow e^{-\Delta t/T_1} + \Gamma_\downarrow ) & T_1(\Gamma_\downarrow e^{-\Delta t/T_1} + \Gamma_\uparrow)
\end{pmatrix} \nonumber
\end{align}

\noindent where 

\begin{align}
    T_1 = \frac{1}{\Gamma_\uparrow + \Gamma_\downarrow},
\end{align}

\noindent is the energy relaxation timescale, $\Gamma_\uparrow$ is the qubit excitation rate, and $\Gamma_\downarrow$ is the qubit relaxation rate.  The transition matrix as defined accounts for both $T_1$ processes: qubit relaxation and excitation.  Leakage effects are not included in this model.

To account for measurement errors we include a confusion matrix $\hat{C}$, which generates the measured probabilities $\vec{p}^{~m}$ from a linear combination of the true probabilities

\begin{align}
    \vec{p}^{~m} &= \hat{C} \vec{p}^{~a},
\end{align}

\begin{align}\label{eqn:confusion_matrix}
\hat{C} &= \begin{pmatrix}
                    p_{00} & p_{10} \\
                    p_{01} & p_{11}
            \end{pmatrix} ,
\end{align}

\noindent where the probabilities $p_{i,j}$ are the probability of state $i$ being measured as state $j$.  

\begin{figure}[tbp]
    \begin{center}
        \includegraphics[width=0.35\columnwidth]{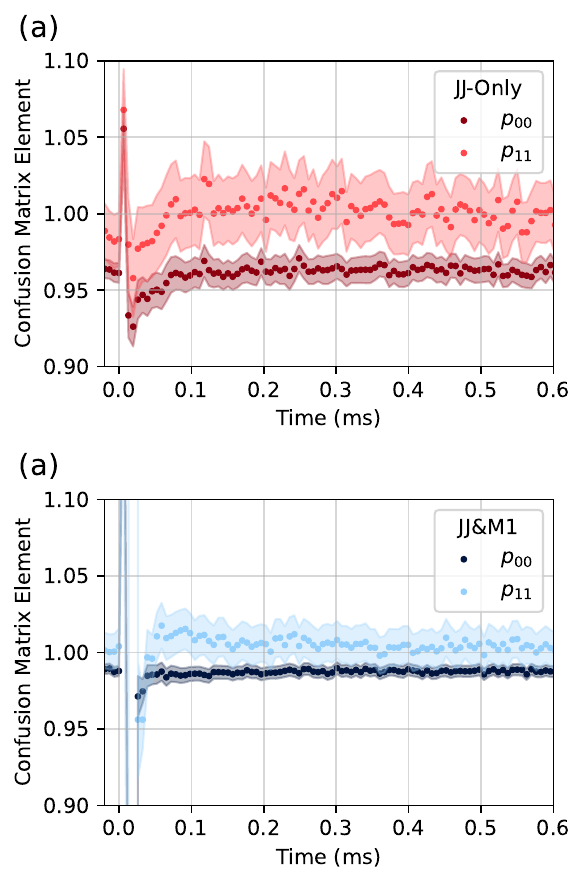}
    \end{center}
    \caption{Confusion matrix parameters $p_{00}$ and $p_{11}$ for the (a) JJ-Only and (b) JJ\&M1 arrays.  The data here results of fits described in Section~\ref{sec:model_main_text}.}
    \label{fig:confusion_matrix}
\end{figure}

We can use the transition matrix to isolate the probability $p_{i\rightarrow j}^a$ of a qubit in state $i$ transitioning to state $j$ after a time $\Delta t$ as:

\begin{align}
    p_{1\rightarrow 0}^a(\Delta t) & = T_1\Gamma_\downarrow\Big(1 - e^{-\Delta t/T_1}\Big), \label{eqn:transition_rates1}\\
    p_{0\rightarrow 1}^a(\Delta t) & = T_1\Gamma_\uparrow\Big(1 - e^{-\Delta t/T_1}\Big). \label{eqn:transition_rates2}
\end{align}

\subsection{Transition Rate Estimation With \Am Data}

In the case of the \Am dataset we do not have sufficient data to constrain the confusion matrix, and therefore rely on Eqns.~\ref{eqn:transition_rates1} and~\ref{eqn:transition_rates2} to extract the transition rates.  We accomplish this by converting our measurements into transition probabilities (described below), and then convert those probabilities into transition rates with Eqns.~\ref{eqn:transition_rates1} and \ref{eqn:transition_rates2}. The relaxation probability is the probability of \e transitioning to \g after \qty{3}{\us} (the end of the X$_\pi$ pulse to the middle of the readout), and the excitation probability is the probability of \g transitioning to \e after \qty{6.95}{\us} (the time between any two subsequent readouts).  Once we have extracted $\Gamma_\uparrow$ and $\Gamma_\downarrow$, the effective qubit temperature is estimated using Eqn.~\ref{eqn:temperature}.  

This process assumes our X$_\pi$ fidelity does not change during the interaction.  With our X$_\pi$ duration of \qty{500}{\ns} the null-to-null spectral width of our pulse is \qty{4}{\MHz}.  From Fig.~\ref{fig:detuning} we observe that the qubit frequencies shift less than \qty{100}{\kHz}, therefore we expect the X$_\pi$ is largely unaffected by qubit frequency shifts due to radiation.

\subsection{Transition Rate Estimation With Linac Data}\label{appendix:linac_rates}

\begin{figure}[tbp]
    \begin{center}
        \includegraphics[width=0.5\columnwidth]{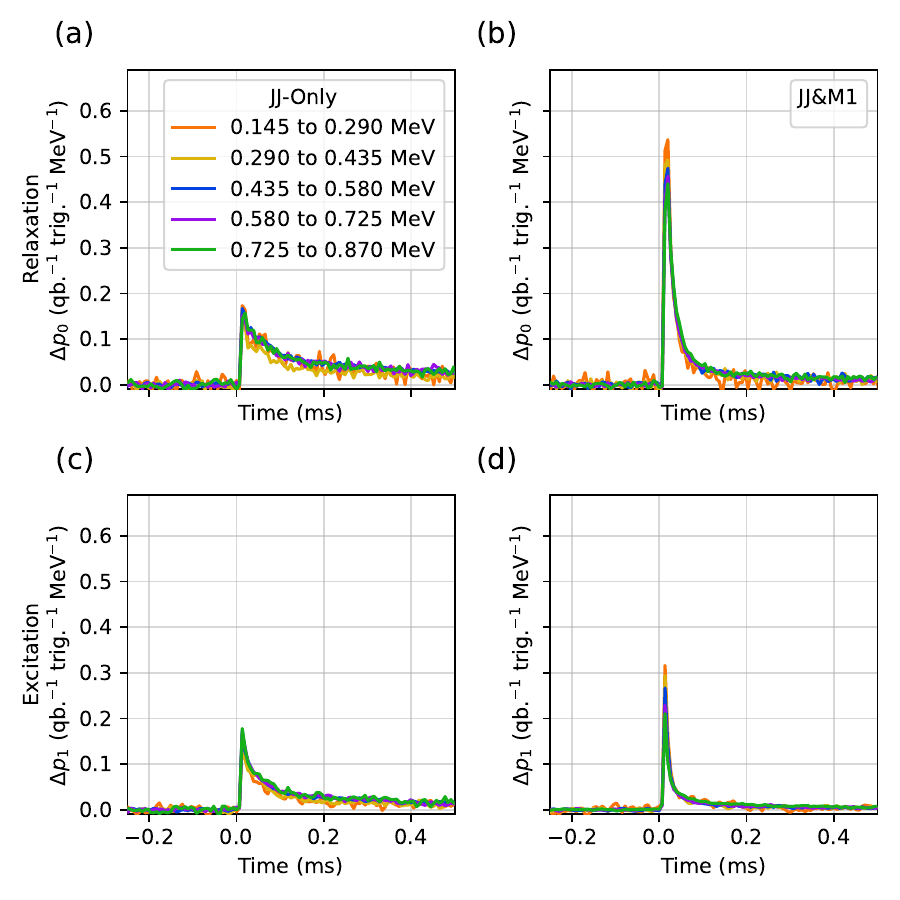}
    \end{center}
    \caption{Energy-normalized response of the (A) and (B) JJ-Only and (C) and (D) JJ\&M1 arrays to the electron linear accelerator as a function of accelerator current.  Traces are the average of approximately~5000 accelerator triggers. (A) and (C) (top row), (B) and (D) (bottom row) show changes in \g and \e state probability, respectively, acting as measures of changes in error rate.}
    \label{fig:energy_linearity}
\end{figure}

\subsubsection{Strategy}

The goal of the following procedure is to convert the measured state probabilities $p_0^m(t)$ and $p_1^m(t)$ into the qubit excitation and relaxation transition rates $\Gamma_\uparrow(t)$ and $\Gamma_\downarrow(t)$.  All told, at any given time point there are four variables that must be solved for: the desired transition rates $\Gamma_\uparrow$ and $\Gamma_\downarrow$, and the confusion matrix parameters $p_{00}$ and $p_{11}$.  The four measurement sequences (A, B, D$_{|0\rangle}$, D$_{|1\rangle}$ in Fig.~\ref{fig:sequences}) yield four data points for each time step after the accelerator trigger, enough to constrain these four parameters of interest.  To accomplish this, we:

\begin{enumerate}
    \item Assume parameters $\Gamma_\uparrow$, $\Gamma_\downarrow$, $p_{00}$ and $p_{11}$,
    \item Evolve the measured probabilities from time $t$ to $t + \Delta t$ using the four measurement sequences and given the assumed parameters,
    \item Perform least-squares minimization between the predicted and measured $\vec{p}(t + \Delta t)$.
\end{enumerate}

\noindent Least-squares minimization was done in \texttt{Python} with \texttt{iminuit}'s \texttt{migrad} algorithm.  

We observed that in measurements shortly after the radiation impact, approximately the first \qty{50}{\micro\s}, the converged values of $\Gamma_\uparrow$, $\Gamma_\downarrow$, $p_{00}$, and $p_{11}$ can take on unphysical values such as transition rates less than zero and probabilities greater than one. Examples of these effects are shown in Figs.~\ref{fig:confusion_matrix} and~\ref{fig:transition_rates_unconstrained}.  We attribute this to the participation of states $\geq |2\rangle$, evidenced by deviations from the $|0\rangle - |1\rangle$ axis in IQ space, Fig.~\ref{fig:iq}.  The participation of these higher states violates the assumption of the two-state measurement sequence model, invalidating the conversion process outlined above.  To work around this constraint, we fix $p_{00}$ and $p_{11}$ to their values estimated \qty{99.6}{\ms} after a particle interaction.  We then obtain $\Gamma_\uparrow$ and $\Gamma_\downarrow$ at all time points and use the fit residual ($\chi^2$) to quantify the validity of assuming $p_{00}$ and $p_{11}$ are constant.  In the plots that follow, we cut data points with $\chi^2 > $ 5.991, a bound which should retain \qty{95}{\percent} of data points that follow a $\chi^2$ distribution with the expected two degrees of freedom.  For completeness, Fig.~\ref{fig:transition_rates_unconstrained} shows the result of allowing $\hat{C}$ to vary in the determination of the rates.

\subsubsection{Matrix Form of Measurement Sequences}

The matrix form of our measurement sequences is described below.  Throughout, $\vec{p}^a(t)$ represents the true qubit state probabilities at time $t$, $t_1$ is the time between the end of the X$_\pi$ pulse and the middle of the measurement, $t_2$ is the time between the middle of the measurement and the beginning of the X$_\pi$ pulse, and $t_3 = t_1 + t_2 + 500~\mathrm{ns}$ is period of a single measurement cycle.  We assume 100\% X$_\pi$ pulse fidelity, and the X$_\pi$ pulse is defined as:

\begin{align}
    \mathrm{X}_\pi = \begin{pmatrix}
                    0 & 1 \\
                    1 & 0
            \end{pmatrix}.
\end{align}

\underline{Sequence A:}

\begin{align}
    \vec{p}^a(t+\Delta t) = \hat{D}(t_1)\mathrm{X}_\pi \hat{D}(t_2) \vec{p}^a(t).
\end{align}

\noindent Here, the measured state at time $t$ is evolved by $\hat{D}(t_2)$, a X$_\pi$ pulse is applied, and then the state is evolved by $\hat{D}(t_1)$ until measurement at time $t+\Delta t$.

\underline{Sequence B:}

\begin{align}
    \vec{p}^a(t+\Delta t) = \hat{D}(t_3) \vec{p}^a(t).
\end{align}

\noindent Here, the measured state at time $t$ is evolved by $\hat{D}(t_3)$  until measurement at time $t+\Delta t$.

\underline{Sequence D$_{|0\rangle}$:}
    
\begin{align}
    \vec{p}^a(t+\Delta t) = 
    &\Big((1 - p_{11})\hat{D}(t_3) + p_{11}\hat{D}(t_1) \mathrm{X}_\pi \hat{D}(t_2)\Big)
    \begin{pmatrix}
        0\\
        p_e^a(t)
    \end{pmatrix}\nonumber\\
    & + \Big(p_{00}\hat{D}(t_3) - (p_{00}-1)\hat{D}(t_1)\mathrm{X}_\pi \hat{D}(t_2)\Big)
    \begin{pmatrix}
        1 - p_e^a(t)\\
        0
    \end{pmatrix}
\end{align}

\noindent We now need to take into account the logic of active reset.  If the qubit is actually in \e and we measure it to be in \e we correctly apply $\hat{D}(t_1) \mathrm{X}_\pi \hat{D}(t_2)$.  If the qubit is actually in \e and we measure it to be in \g we mistakenly apply $\hat{D}(t_3)$.  If the qubit is actually in \g and we measure it to be in \e we mistakenly apply $\hat{D}(t_1) \mathrm{X}_\pi \hat{D}(t_2)$. If the qubit is actually in \g and we measure it to be in \g we correctly apply $\hat{D}(t_3)$.

\underline{Sequence D$_{|1\rangle}$:}
    
\begin{align}
    \vec{p}^a(t+\Delta t) =
    & \Big((1 - p_{11})\hat{D}(t_1)\mathrm{X}_\pi \hat{D}(t_2) - p_{11}\hat{D}(t_3)\Big)
    \begin{pmatrix}
        0\\
        p_e^a(t)
    \end{pmatrix}\nonumber\\
    + &\Big(p_{00}\hat{D}(t_1)\mathrm{X}_\pi \hat{D}(t_2) + (1 - p_{00})\hat{D}(t_3)\Big)
    \begin{pmatrix}
        1 - p_e^a(t)\\
        0
    \end{pmatrix}
\end{align}

\noindent In this case, if the qubit is actually in \e and we measure it to be in \e we correctly apply $\hat{D}(t_3)$.  If the qubit is actually in \e and we measure it to be in \g we mistakenly apply $\hat{D}(t_1) \mathrm{X}_\pi \hat{D}(t_2)$.  If the qubit is actually in \g and we measure it to be in \e we mistakenly apply $\hat{D}(t_3)$. If the qubit is actually in \g and we measure it to be in \g we correctly apply $\hat{D}(t_1) \mathrm{X}_\pi \hat{D}(t_2)$.

\subsubsection{Generating Probability Traces and Quality Cuts}

The raw probability data for this calculation are constructed as follows.  We define a ``trace'' as the average qubit response during the course of one measurement run.  The data to be converted into rates is the average of traces with the estimated mean electron flux between one and six electrons, and where the average probability in the first 5000 samples is not more than $2\sigma$ different than the average probability in the last 5000 samples.  This quality cut excludes runs where the qubit state dramatically changed during the course of the run.  In the following, we rely on the assumption that our qubit response $\Delta p$ is linear in the energy deposited into the Si substrate.  This assumption incurs, on average, an approximately $9\%$ systematic uncertainty on the recovery timescale, estimated from the spread exponential decay times fitting to normalized average traces at a variety of different energies.  This weighs lower energy impacts slightly more than higher energy impacts [Fig.~\ref{fig:energy_linearity}].  We accept this degree of uncertainty given the qualitative nature of our comparisons.  To determine the average response for each qubit, and to obtain a response which can be compared across qubits, we perform the following procedure:

\begin{enumerate}
    \item Cut traces based on the quality metrics described above.
    \item Compute the average baseline, weighted by the number of triggers in the run.
    \item Baseline subtract each trace, with the baseline defined as the mean of the first 40 data points in the trace.  This results in traces in $\Delta p$.
    \item Compute the average of traces, weighted by the number of triggers in the run.
    \item Using the measured linac beam currents, compute the average energy deposited into the silicon, weighted by the number of triggers in each run and the uncertainty on each run's energy deposited measurement uncertainty.
    \item Divide the trace by the average energy deposited, resulting in $\Delta p$-per-keV.
    \item Multiply by \qty{508}{\keV} to convert the trace to $\Delta p$-per-\qty{508}{\keV} (\qty{3.5}{\electron}), the center of the energy deposited window considered.  This choice attempts to minimize the effect of the $9\%$~nonlinearity.
    \item As the transition rate estimation relies on absolute probabilities, add the average baseline to the average $\Delta p$ trace and return.
\end{enumerate}

\noindent The above procedure results in a single, average trace of $p_0$ for each measurement sequence and qubit in each array, resulting in 72~ total traces (18~qubits and 4~measurement~sequences).  The measured and modeled transition rates are consistent with the \Am dataset [Fig.~\ref{fig:alphas}]: the steady-state rates are similar, and the transition rates at the event peak are smaller than observed in the \Am dataset due to the lower incident energies. 

\begin{figure}[tbp]
    \begin{center}
    \includegraphics[width=0.5\columnwidth]{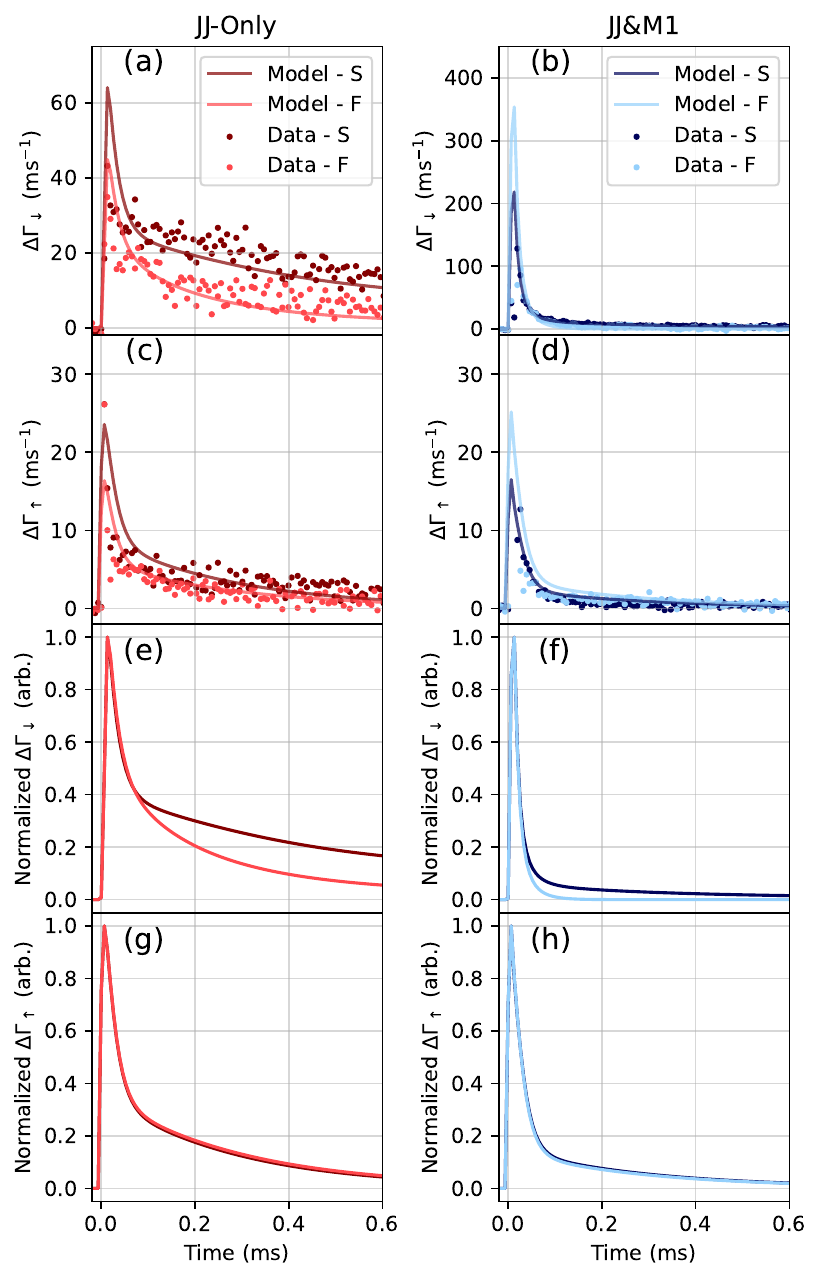}
    \end{center}
    \caption{Average transition rates (dots) and comparison with the QP model (lines).  We show the average change in relaxation and excitation rate for the JJ-Only array in panels (a), (c), (e), and (g).  Dark red and light red represent slow and fast JJ orientations respectively.    We show the average change in relaxation and excitation rate for the JJ-Only array in panels (b), (d), (f), and (h).  Dark blue and light blue represent represent slow and fast JJ orientations respectively.  Neutral blue shows the array average.  Panels (e)-(h) show the model from (a)-(d), except normalized to a peak value of one for comparison with Fig.~\ref{fig:orientation_and_device}.  Panels (a), (b), (e), and (f) show the qubit relaxation rate, while panels (c), (d), (g), and (h) show the qubit excitation rate.  In comparison with Fig.~\ref{fig:transition_rates}, the confusion matrix parameters are not held constant at any point.}
    \label{fig:transition_rates_unconstrained}
\end{figure}

\section{Quasiparticle Model and Comparison with Data}\label{sec:modeling_appendix}

\subsection{Model}\label{sec:modeling_appendix_density}

The model described in Sec.~\ref{sec:model_main_text} is written in full below.  At stated in the main text, the model is a system of six differential equations: one for the probability that the qubit occupies \g, $p_0$, and five for the following QP densities

\begin{align}
    \Big\{x_i~\Big|~i \in \{2<,~2>,~3,~L,~R\}\Big\}.
\end{align}

\noindent The QP density subscripts are described in Fig.~\ref{fig:devices} and Sec.~\ref{sec:modeling}.  The base structure of the equations for QP density is in the style of the ``Rothwarf-Taylor'' phenomenological model, Eqn.~\ref{eqn:RT}.  Below, we define a ``lead'' as a JJ-contacting metal layer (one of either M2 or M3).

The equations describing the M1 material are exclusively in the form of Eqn.~\ref{eqn:RT}.  The generation term, Eqn.~\ref{eqn:generation}, is identical to that given in \MG, and the term with linear trapping is adapted as $s_ix_i\rightarrow s_i(x_i - x_{i,0})$ such that the QP density at equilibrium is $x_{i,0}$.  Throughout, we assume that $x_{i,0} = 10^{-8}$~\cite{connollyCoexistenceNonequilibriumDensity2024a}.  We hypothesize that our observed JJ orientation dependence is due to a difference in the effective trapping terms $s_i$ between the qubit capacitor and the ground plane.  Accordingly we set $s_{L}\neq s_{R}$.  Qubits with different JJ orientations are modeled by switching the role of $s_{L} \leftrightarrow s_{R}$, changing the behavior of the QP density coupled to layer M2.  The trapping term associated with the qubit capacitor is assumed to be $s = \qty{500}{\per\s}$, and the ground plane is assumed to be $s = \qty{50e3}{\per\s}$.  These values are consistent with previous literature~\cite{yeltonModelingPhononmediatedQuasiparticle2024,diamondDistinguishingParitySwitchingMechanisms2022,wangMeasurementControlQuasiparticle2014}.

In the equations describing the layers M2 and M3 ($x_{2<}$, $x_{2>}$, $x_{3}$) we largely follow \MG.  These start in the form of Eqn.~\ref{eqn:RT}, dropping the linear trapping term, and incorporate terms coupling the QP densities together through qubit transitions and QP tunneling between films.  The equations for $x_{2<}$ and $x_{2>}$ are further coupled by a temperature-dependent characteristic QP excitation and relaxation timescale.  The following are changes relative to \MG:

\begin{enumerate}
    \item In \MG the generation term combines photon assisted tunneling and pair-breaking phonons --- here we drop terms related to photon assisted tunneling, assuming it will be subdominant.  The generation term therefore is strictly from pair-breaking phonons, tied to the assumed temperature $T$.
    \item As the layers M2 and M3 in this work have different thicknesses, we incorporate a term $\nu$ accounting for the volume difference between films $2$ and $3$.  
    \item To incorporate the dependence on \ddgnd and JJ orientation we add terms which couple M2 and M3 to the M1 films, $x_{L}$ and $x_{R}$.  
\end{enumerate}

QPs enter the layers M2 and M3 through a term proportional to a product of $x_{j}^{>\Delta_i}(T)$, the QP density in the M1 material $j$ adjacent lead $i$ with energy greater than the superconducting gap of lead $i$, $\nu_s$, an overall scale factor accounting for the difference in volume between the M1 and both M2 and M3, and one of either $\nu_{tk}$ or $\nu_{tn}$, a lead-specific scale factor accounting for the effective thickness of layer M2 and M3 respectively.  The products $\nu_s\nu_{tn}$ and $\nu_s\nu_{tk}$ are greater than $10^3$, which we interpret as the effective volume difference between layer M1 and the adjacent JJ-contacting layer (M2 or M3).  Agreement between the model and data is best when  $\nu_s\nu_{tk}$ is approximately $4.4\times$ greater than what would be expected from the M2/M3 layer thickness.  We hypothesize this could be due to a contribution from QP generation in the unmodeled M3 layer on the low-gap side of the JJ --- QPs generated in this layer would fall into the lower gap layer M2.  This would most strongly affect layer M2 in the JJ-Only array due to the larger \ddjj, meaning layer M2 is acting as a stronger trap for QPs produced in this unmodeled layer.

QPs drain from layers M2 and M3 into layer M1 through a term proportional to the product of $x_i$, the QP density in lead $i$, a scale factor $\eta_i$ unique to lead $i$ but independent of lead thickness or superconducting gap, and $\tau_{x_i}^{-1}(T)$, a temperature-dependent QP relaxation timescale related to \ddgnd.  The model agrees with the data best with $\eta_3=4000$ and $\eta_2=115$.  We interpret $\eta_3>\eta_2$ to be as a result of layer M3's direct contact with layer M1, a consequence of the Dolan bridge style JJs.  The agreement of this QP model with data, despite the fact that $\eta_i\gg0$, suggests that this phenomenology can be refined in future work.  Additionally, we do not account for the effects of QP diffusion.  As the goal of this exercise is to demonstrate qualitative agreement with the data we leave these refinements to future work.

With inspiration from \MG, we define $\tau_{x_i}^{-1}(T)$ as:

\begin{align}\label{eqn:tau_inv}
    \tau_{x_i}^{-1}(T) = A(\Delta_i,T)\int_{\varepsilon_\text{min}}^{\varepsilon_\text{max}}d\varepsilon\int_{\omega_\text{min}(\varepsilon)}^{\omega_\text{max}(\varepsilon)}d\omega N(\Delta_i,\varepsilon)e^{-(\varepsilon - \Delta_i) / k_\text{B}T}\frac{1}{\rho_{\text{qp},i}}\omega^2
\left(1+n(\omega,T)\right)
\frac{\varepsilon(\varepsilon-\omega)-\Delta_1^2}
{\varepsilon\sqrt{(\varepsilon-\omega)^2-\Delta_1^2}},
\end{align}

\noindent Where

\begin{align}
    A(\Delta_i,T) = \frac{2\sqrt{2\pi}}{\sqrt{\Delta_i k_\text{B}T}}\frac{r}{8\pi\Delta_i},
\end{align}

\noindent Is a constant factor,

\begin{align}
    N(\Delta_i,\varepsilon) = \frac{\varepsilon}{\varepsilon^2 - \Delta_i^2},
\end{align}

\noindent Is the superconducting density of states,

\begin{align}
    \rho_{\text{qp},i} = 
    \begin{cases}
    1 & (i = 3) \\
    \operatorname{erfc}\!\left(
\sqrt{\dfrac{\delta\Delta_\text{JJ}}{k_\text{B} T}}
\right)
& (i = 2>) \\
    \operatorname{erf}\!\left(
\sqrt{\dfrac{\delta\Delta_\text{JJ}}{k_\text{B} T}}
\right)
& (i = 2<), \\
    \end{cases}
\end{align}

\noindent Is a factor accounting for the QP density partitioning between $2<$ and $2>$, and the Bose-Einstein distribution is

\begin{align}
    n(\omega,T)=\frac{1}{e^{\omega/(k_BT)}-1}.
\end{align}

The integration limits are given as follows.  Below, in cases $i = 3, 2>$, the limits for $\varepsilon_{\max}$  are chosen for numerical convenience and are approximately infinite.  Integration is performed in \texttt{Python} using \texttt{scipy.integrate.dblquad}.  To overcome numerical limitations of the algorithm, the integration over $\varepsilon$ for $i=3$ is broken into \qty{2000} evenly spaced steps and then summed.  For $i = 3$:

\begin{align*}
\varepsilon_{\min}&=\Delta_3,\\
\varepsilon_{\max}&=2\Delta_3,\\
\omega_{\min}(\varepsilon)&=\varepsilon-\Delta_3,\\
\omega_{\max}(\varepsilon)&=\varepsilon-\Delta_1.
\end{align*}

\noindent For $i = 2>$

\begin{align*}
\varepsilon_{\min}&=\Delta_3,\\
\varepsilon_{\max}&=4\Delta_3,\\
\omega_{\min}(\varepsilon)&=\varepsilon-\Delta_3,\\
\omega_{\max}(\varepsilon)&=\varepsilon-\Delta_1.
\end{align*}

\noindent For $i = 2<$

\begin{align*}
\varepsilon_{\min}&=\Delta_2,\\
\varepsilon_{\max}&=\Delta_3,\\
\omega_{\min}(\varepsilon)&=\varepsilon-\Delta_2,\\
\omega_{\max}(\varepsilon)&=\varepsilon-\Delta_1.
\end{align*}

The system of differential equations described above is presented below:

\begin{align}
\dot{x}_3 &= g(T, r, \Delta + \delta\Delta_\text{JJ}) - r x_3^2 \label{eqn:x3}\\
 &\quad - \delta \Big[ (\bar{\Gamma}_{00}^3(T) + \bar{\Gamma}_{01}^3(T)) p_0 + (\bar{\Gamma}_{11}^3(T) + \bar{\Gamma}_{10}^3(T)) p_1 \Big] x_3 \tilde{\bar{s}} \nonumber\\
&\quad + \delta\, \bar{\Gamma}_{10}^{2<}(T) p_1 x_{2<} \tilde{\bar{s}} \nu 
 + \delta \Big[ (\bar{\Gamma}_{00}^{2>}(T) + \bar{\Gamma}_{01}^{2>}(T)) p_0 + (\bar{\Gamma}_{11}^{2>}(T) + \bar{\Gamma}_{10}^{2>}(T)) p_1 \Big] x_{2>} \tilde{\bar{s}} \nu \nonumber\\
&\quad + \nu_{th} \nu_s x_{L}^{>\Delta_3}(T) 
- \eta_3\, x_3\, \tau_{x3}^{-1}(T) \nonumber
\\[12pt]
\dot{x}_{2>} &= g_>(T, r, \Delta)
 - r x_{2>}^2 
 - r x_{2>} x_{2<} \\
&\quad - \Big[ (\bar{\Gamma}_{00}^{2>}(T) + \bar{\Gamma}_{01}^{2>}(T)) p_0 + (\bar{\Gamma}_{11}^{2>}(T) + \bar{\Gamma}_{10}^{2>}(T)) p_1 \Big] x_{2>} \tilde{\bar{s}} \nonumber\\
&\quad + \Big[ \bar{\Gamma}_{00}^3(T) p_0 + (\bar{\Gamma}_{11}^3(T) + \bar{\Gamma}_{10}^3(T)) p_1 \Big] x_3 \frac{\tilde{\bar{s}}}{\nu} + \xi(T) \bar{\Gamma}_{01}^3(T) p_0 x_3 \frac{\tilde{\bar{s}}}{\nu}
  \nonumber\\
&\quad - \tau_\text{Rlx.}^{-1}(T)  x_{2>}
 + \tau_\text{Exc.}^{-1}(T)  x_{2<} \nonumber\\
&\quad + \nu_{tk} \nu_s x_{R}^{>\Delta_3}(T) 
 - \eta_0\, x_{2>}\, \tau_{x2>}^{-1}(T) \nonumber
\\[12pt]
\dot{x}_{2<} &= g_<(T, r, \Delta)
 - r x_{2<}^2 
 - r x_{2>} x_{2<}\\
&\quad - \bar{\Gamma}_{10}^{2<}(T) p_1 x_{2<} \tilde{\bar{s}} + \Big[1 - \xi(T)\Big] \bar{\Gamma}_{01}^3(T) p_0 x_3 \frac{\tilde{\bar{s}}}{\nu}
  \nonumber\\
&\quad + \tau_R^{-1}(T)  x_{2>}
 - \tau_E^{-1}(T)  x_{2<} \nonumber\\
&\quad + \nu_{tk} \nu_s \Big[x_{R}^{\Delta_0}(T) - x_{R}^{\Delta_3}(T)\Big]
 - \eta_0\, x_{2<}\, \tau_{x2<}^{-1}(T) \nonumber
\\[12pt]
\dot{x}_{L} &= g(T, r, \Delta - \delta\Delta_\text{M1})
 - s_{L} (x_{L} - x_{L,0})
 - r x_{L}^2
\\[12pt]
\dot{x}_{R} &= g(T, r, \Delta - \delta\Delta_\text{M1})
 - s_{R} (x_{R} - x_{R,0})
 - r x_{R}^2
 \\[12pt]
\dot{p}_0 &= -\Big[ \Gamma_{01}^{eo} + \Gamma_{01}^{ee}(T)\, \Big] p_0 
 + \Big[ \Gamma_{10}^{eo} + \Gamma_{10}^{ee}\, \Big] p_1
\end{align}

\noindent Where,

\begin{align}
\Gamma_{01}^{eo} &\equiv \frac{1}{\tilde{\bar{s}}} \Big[ \tilde{\Gamma}_{01}^3(T) x_3 + \tilde{\Gamma}_{01}^{2>}(T) x_{2>} \Big]
\\
\Gamma_{10}^{eo} &\equiv \frac{1}{\tilde{\bar{s}}} \Big[ \tilde{\Gamma}_{10}^3(T) x_3 + \tilde{\Gamma}_{10}^{2>}(T) x_{2>} + \tilde{\Gamma}_{10}^{2<}(T) x_{2<} \Big]
\\[8pt]
\Gamma_\uparrow &\equiv \Gamma_{01}^{eo} 
+ \Gamma_{01}^{ee}(T)\, \\[6pt]
\Gamma_\downarrow &\equiv \Gamma_{10}^{eo} 
+ \Gamma_{10}^{ee}\,\label{eqn:relaxation}\\[6pt]
x_{j}^{>\Delta_i}(T) &\equiv x_{j} \text{erfc}\Bigg(\sqrt{\frac{\Delta_i - \Delta_{j}}{k_B T}}\Bigg)
\end{align}

\noindent Here, a dot denotes the derivative with respect to time, and all variables are defined in Table~\ref{tab:model_vars}.  

\begin{table*}
    \centering
    \begin{tabular}{l l}
        \toprule
        Variable & Definition \\
        \midrule
        $x_i$ & The reduced QP density in superconducting film $i$. \\
        $g$ & QP generation term, defined in Eqn.~\ref{eqn:generation}. \\
        $g_>,~g_<$ & QP generation term for QPs greater than and less than $\Delta$, respectively.  \\
        & Defined in Eqns.~\ref{eqn:generation_conditional<} and \ref{eqn:generation_conditional>}.  \\
        $T$ & Effective temperature of the quasiparticle/phonon system.  A parametrization, does not strictly correspond\\
            & to the system's physical temperature. \\
        $r$ & Quadratic ``recombination'' constant. \\
        $\Delta$ & Superconducting gap in layer M2. \\
        $\delta\Delta_\text{JJ}$  & Difference in superconducting gap between layers M2 and M3.  \\
        $\delta$   &  Ratio of superconducting gap between layers M2 and M3, defined in Eqn.~\ref{eqn:delta}. \\
        $\bar{\Gamma}_{if}^{\alpha}$  & Tunneling rate for a single QP in electrode $\alpha\in\{2<,~2>,~3,~L,~R\}$ from qubit state $i$ to qubit state $f$. \\
          & Defined in \MG. \\
        $p_i$ & Probability of finding the qubit in state $i\in\{0,~1\}$.\\
        $\tilde{\bar{s}}$   & Overall volume scaling factor.  Relative to the default value of 3400.32~$\mu$m$^3$ in \MG.  \\
        $\nu$   & Volume of the low gap (thick) lead divided by the volume of the high gap (thin) lead.  \\
        $\nu_{th}$ & Thickness of M1 divided by the thickness of M3.\\
        $\nu_{tk}$ & Thickness of M1 divided by the effective thickness of M2.  Effective thickness in Table~\ref{tab:model_params}.\\
        $\nu_s$ & Scaling factor accounting for the volume difference between layer M1 and the sum of M2 and M3.\\
        $\eta_i$ & Scale factor for QP relaxation in film $i\in\{2,~3\}$.\\
        $\tau_{x_i}^{-1}(T)$ & QP relaxation timescale from film $i\in\{2<,~2>,~3\}$ to layer M1.  Defined in Eqn.~\ref{eqn:tau_inv}.\\
        $\xi(T)$ & Dimensionless factor denoting the fraction of QPs tunneling from M3 to M2 with final \\
          & energy greater than $\Delta$ during a qubit excitation process.  Defined in \MG. \\
        $\tau_\text{Rlx.}^{-1}$(T) & QP relaxation timescale in layer M2 from $E_{qp}>\Delta + \delta\Delta_\text{JJ}$ to $E_{qp}\in\big[\Delta,\Delta + \delta\Delta_\text{JJ}\big]$.  Equivalent\\
            & to $\tau_R^{-1}$ in \MG.\\
        $\tau_\text{Exc.}^{-1}$(T) & QP excitation timescale in layer M2 from $E_{qp}\in\big[\Delta,\Delta + \delta\Delta_\text{JJ}\big]$ to $E_\text{qp}> \Delta + \delta\Delta_\text{JJ}$.  Equivalent\\
            & to $\tau_E^{-1}$ in \MG.\\
        $\delta\Delta_\text{M1}$ & Superconducting gap difference between layers M2 and M1. \\
        $s_i$ & Effective linear trapping parameter in film $i\in\{L,~R\}$.\\
        $\Gamma_{01}^{eo}$ & Total parity-switching qubit excitation rate. \\
        $\Gamma_{01}^{ee}(T)$ & Total non-parity-switching qubit excitation rate.  Defined in Eqn.~\ref{eqn:rlx_to_exc}. \\
        $\Gamma_{10}^{eo}$ & Total parity-switching qubit relaxation rate. \\
        $\Gamma_{01}^{ee}$ & Total non-parity-switching qubit relaxation rate. \\
        $\tilde{\Gamma}_{if}^{\alpha}$   & Per-QP tunneling rate ($\bar{\Gamma}_{if}^{\alpha}$) times the number of Cooper pairs in layer M2.  \\
            & Defined in \MG. \\
        $\Gamma_\uparrow$ & Total qubit excitation rate. \\
        $\Gamma_\downarrow$ & Total qubit relaxation rate. \\
        $x_{j}^{>\Delta_i}(T)$ & QP density in layer M1 with energy greater than $\Delta_i$ with $j\in{L,R}$ and $i\in{2,3}$.\\
        $T_\text{scale,fall1}$     & Defined in Eqn.~\ref{eqn:temp}.\\
        $T_\text{scale,fall2}$  & Defined in Eqn.~\ref{eqn:temp}.\\
        $\tau_\text{fall1}$     & Defined in Eqn.~\ref{eqn:temp}.\\
        $\tau_\text{fall2}$     & Defined in Eqn.~\ref{eqn:temp}.\\
        $\tau_\text{rise1}$     & Defined in Eqn.~\ref{eqn:temp}.\\
        $N_i$ & Number of QPs in film $i$. \\
        $n_{\text{CP},i}$ & Cooper pair density in film $i$, defined in Eqn.~\ref{eqn:ncp}.\\
        $V_i$ & Volume of film $i$.\\
        $N_{j}^{>\Delta_i}$ & Number of QPs in layer M1 with energy greater than $\Delta_i$ with $j\in{L,R}$ and $i\in{2,3}$.\\
        \bottomrule
    \end{tabular}
    \caption{Symbol and variable definitions for the QP model described in Eqns.~\ref{eqn:x3}-\ref{eqn:relaxation}.}
    \label{tab:model_vars}
\end{table*}

The following definitions are referenced in Table~\ref{tab:model_vars}.  The QP generation term from thermal phonons is copied from \MG:

\begin{align}\label{eqn:generation}
    g(T, r, \Delta) = \frac{1}{\Delta}
  2\pi r \, k_B T \,
  e^{-\dfrac{2\Delta}{k_B T}},
\end{align}

\noindent where $k_B$ is Boltzmann's constant.  The variants:

\begin{align}\label{eqn:generation_conditional<}
    g_<(T, r, \Delta) = g(T, r, \Delta) \operatorname{erf}\!\left(
    \sqrt{\frac{\delta\Delta_\text{JJ}}{k_B T}}
\right),\\\label{eqn:generation_conditional>}
    g_>(T, r, \Delta) = g(T, r, \Delta) \operatorname{erfc}\!\left(
    \sqrt{\frac{\delta\Delta_\text{JJ}}{k_B T}}
\right),
\end{align}

\noindent account for QP generation below and above $E_\text{qp}= \Delta + \delta\Delta_\text{JJ}$, respectively.  The functions ``erf'' and ``erfc'' are the error function and complementary error function respectively.  The scaling factor $\delta$ is defined as:

\begin{align}\label{eqn:delta}
    \delta = \frac{\Delta}{\Delta + \delta\Delta_\text{JJ}}.
\end{align}

\noindent The total non-parity-switching qubit excitation rate is related to the non-parity-switching qubit relaxation rate through a Boltzmann factor:

\begin{align}\label{eqn:rlx_to_exc}
    \Gamma_{01}^{ee}(T) = \Gamma_{10}^{ee}e^{-hf_\text{qb}/k_BT},
\end{align}

\noindent where $h$ is Planck's constant and $f_\text{qb}$ is the qubit frequency.

The time-dependent temperature is parameterized as
    
\begin{align}\label{eqn:temp}
T(t) = T_b
+ T_{\text{scale}}\, A\,
\left(
T_{\text{scale,fall1}} \, e^{-\frac{t - t_0}{\tau_{\text{fall1}}}}
+
T_{\text{scale,fall2}} \, e^{-\frac{t - t_0}{\tau_{\text{fall2}}}}
\right)
\left( 1 - e^{-\frac{t - t_0}{\tau_{\text{rise}}}} \right)
\end{align}

\noindent where

\begin{align}
A &= \frac{\qty{0.33}{\K} - T_b}{\Delta T},\label{eqn:A}\\
    \Delta T &= 
\max \Bigg\{
\left(
T_{\text{scale,fall1}} \, e^{-\frac{t - t_0}{\tau_{\text{fall1}}}}
+
T_{\text{scale,fall2}} \, e^{-\frac{t - t_0}{\tau_{\text{fall2}}}}
\right)
\left( 1 - e^{-\frac{t - t_0}{\tau_{\text{rise}}}} \right)
\Bigg\}\label{eqn:deltaT}
\end{align}

\noindent are scale factors defined for the convenience of matching the model to data. 

The Cooper pair density is defined as:

\begin{align}\label{eqn:ncp}
    n_{cp,i} = 2\nu_i\Delta_i,
\end{align}

\noindent where $\nu_i$ is the single-spin density of states in film $i$.

The key parameters for the QP model are shown in Table~\ref{tab:model_params}.

\begin{table}[htbp]
    \centering
    \begin{tabular}{l l}
    \toprule
    Parameter & Value \\
    \midrule
    Capacitor trapping $s$     & $500~\text{s}^{-1}$ \\
    Ground plane trapping $s$     & $50\times10^{3}~\text{s}^{-1}$ \\
    Recombination $r$       & $7\times10^{6}~\text{s}^{-1}$ \\
    $\tilde{\bar{s}}$     & $3\times10^{3}$ \\
    $\nu_s$     & $750$ \\
    $\nu_{tk}$ JJ\&M1 & 250~nm / 30~nm\\
    $\nu_{tk}$ JJ-Only & 250~nm / 45~nm\\
    $\eta_0$    & $115$ \\
    $\eta_3$    & $4000$\\
    $T_\text{scale,fall1}$     & 1\\
    $t_0$   & 0\\
    $T_\text{scale,fall2}$ JJ\&M1    & 0.1\\
    $T_\text{scale,fall2}$ JJ-Only    & 0.25\\
    $\tau_\text{fall1}$     & $20$~$\mu$s\\
    $\tau_\text{fall2}$     & $350$~$\mu$s\\
    $\tau_\text{rise1}$     & $5$~$\mu$s\\
    $x_{i,0}$          & $10^{-8}$\\
    \bottomrule
    \end{tabular}
    \caption{Key parameters used in the QP model.  Trapping rates apply to either $s_L$ or $s_R$, depending on the modeled qubit and its JJ orientation.}
    \label{tab:model_params}
\end{table}

\subsection{Quasiparticle Number vs. Density}

It can be illustrative to represent the equations in terms of QP number instead of QP density.  In particular this gives intuition for the influence of different layer thicknesses.  Below, we have performed the substitution:

\begin{align}
    x_i = \frac{N_i}{n_{\text{CP},i}V_i},
\end{align}

\noindent where variables are defined in Table~\ref{tab:model_vars}.  We have introduced two new variables, $h_0$ and $h_{0,\text{eff}}$, as the designed and effective thicknesses of layer M2, respectively.

\begin{align}
\dot{N}_3 &= n_{cp,3}V_3g(T, r, \Delta + \delta\Delta_\text{JJ}) - r N_3^2 /(n_{cp,3}V_3)\label{eqn:x3}\\
 &\quad - \delta \Big[ (\bar{\Gamma}_{00}^3(T) + \bar{\Gamma}_{01}^3(T)) p_0 + (\bar{\Gamma}_{11}^3(T) + \bar{\Gamma}_{10}^3(T)) p_1 \Big] N_3 \tilde{\bar{s}} \nonumber\\
&\quad + \, \bar{\Gamma}_{10}^{2<}(T) p_1 N_{2<} \tilde{\bar{s}} 
     + \Big[ (\bar{\Gamma}_{00}^{2>}(T) + \bar{\Gamma}_{01}^{2>}(T)) p_0 + (\bar{\Gamma}_{11}^{2>}(T) + \bar{\Gamma}_{10}^{2>}(T)) p_1 \Big] N_{2>}\tilde{\bar{s}}  \nonumber\\
&\quad + \frac{\Delta_3}{\Delta_{L}}\nu_s N_{L}^{>\Delta_3} 
 - \eta_3\, N_3\, \tau_{x3}^{-1}(T) \nonumber
\\[12pt]
\dot{N}_{2>} &= n_{cp,0}V_0g_>(T, r, \Delta)
 - r N_{2>}^2 / (n_{cp,0}V_0)
 - r N_{2>} N_{2<} / (n_{cp,0}V_0) \\
&\quad - \Big[ (\bar{\Gamma}_{00}^{2>}(T) + \bar{\Gamma}_{01}^{2>}(T)) p_0 + (\bar{\Gamma}_{11}^{2>}(T) + \bar{\Gamma}_{10}^{2>}(T)) p_1 \Big] N_{2>} \tilde{\bar{s}} \nonumber\\
&\quad + \delta \Big[ \bar{\Gamma}_{00}^3(T) p_0 + (\bar{\Gamma}_{11}^3(T) + \bar{\Gamma}_{10}^3(T)) p_1 \Big] N_3 \tilde{\bar{s}} + \delta \xi(T) \bar{\Gamma}_{01}^3(T) p_0 N_3 \tilde{\bar{s}}
  \nonumber\\
&\quad - \tau_\text{Rlx.}^{-1}(T)  N_{2>}
 + \tau_\text{Exc.}^{-1}(T)  N_{2<} \nonumber\\
&\quad + \frac{\Delta_0}{\Delta_{R}}\frac{h_0}{h_{0,\text{eff}}}\nu_s N_{R}^{>\Delta_3}
 - \eta_0\, N_{2>}\, \tau_{x2>}^{-1}(T) \nonumber
\\[12pt]
\dot{N}_{2<} &= n_{cp,0}V_0g_<(T, r, \Delta)
 - r N_{2<}^2 / (n_{cp,0}V_0)
 - r N_{2>} N_{2<} / (n_{cp,0}V_0)\\
&\quad - \bar{\Gamma}_{10}^{2<}(T) p_1 N_{2<} \tilde{\bar{s}} + \delta (1 - \xi(T)) \bar{\Gamma}_{01}^3(T) p_0 N_3 \tilde{\bar{s}}
  \nonumber\\
&\quad + \tau_\text{Rlx.}^{-1}(T)  N_{2>}
 - \tau_\text{Exc.}^{-1}(T)  N_{2<} \nonumber\\
&\quad + \frac{\Delta_0}{\Delta_{R}}\frac{h_0}{h_{0,\text{eff}}}\nu_s (N_{R}^{>\Delta_0} - N_{R}^{>\Delta_3})
 - \eta_0\, N_{2<}\, \tau_{x2<}^{-1}(T) \nonumber
\\[12pt]
\dot{N}_{L}&= n_{cp}V_{L} g(T, r, \Delta - \delta\Delta_\text{M1})
 - s_{L} (N_{L} - N_{L,0})
 - r N_{L}^2 / (n_{cp}V_{L})
\\[12pt]
\dot{N}_{R} &= n_{cp}V_{R}g(T, r, \Delta - \delta\Delta_\text{M1})
 - s_{R} (N_{R} - N_{R,0})
 - r N_{R}^2 / (n_{cp}V_{R})
 \\[12pt]
\dot{p}_0 &= -\Big[ \Gamma_{01}^{eo} + \Gamma_{01}^{ee}(T)\, \Big] p_0 
 + \Big[ \Gamma_{10}^{eo} + \Gamma_{10}^{ee}\, \Big] p_1
\end{align}

\subsection{Model and Data Comparison}\label{appendix:model_and_data}

We now compare the model in Appendix~\ref{sec:modeling_appendix_density} to the data.  The majority of parameters are fixed across all qubits and arrays, see Table~\ref{tab:model_params}.  Because of the under-constrained nature of the model, these fixed parameters were chosen from physically-motivated ranges by trial and error.  The following parameters, defined in Table~\ref{tab:model_vars}, are exceptions and allowed to vary between all qubits:

\begin{enumerate}
    \item The temperature scale $T_\text{scale}$,
    \item The non parity switching relaxation rate $\Gamma_{10}^{ee}$, which is set to a value within \qty{20}{\percent} of the baseline relaxation rate (the average of the first 50 points), and
    \item Base temperature $T_b$, set to a value within $2\sigma$ of the average baseline temperature estimated with detailed balance, where $\sigma$ is the standard deviation of the first 50 points
\end{enumerate}

\noindent Additionally, the amplitude of the slow component of the temperature fall time ($T_\text{scale, fall2}$) takes a different value for each gap profile/array.  This is justified by the different temperature profiles observed in Fig.~\ref{fig:temperature}.  Specific parameters used are listed in Table~\ref{tab:model_params}.

As a supplement to the discussion from Sec.~\ref{sec:model_main_text}, Fig.~\ref{fig:modeling_by_qubit} breaks the averaged fits and data presented in Fig~\ref{fig:transition_rates} into individual qubits.  We retain qualitative agreement between the data and model output with a few exceptions such as ``JJ-Only - Q1'' and ``JJ-Only - Q4''.  Modeling of these qubits may be confounded by their large baseline transition rates.  

\begin{figure*}[tbp]
        \includegraphics[width=\textwidth]{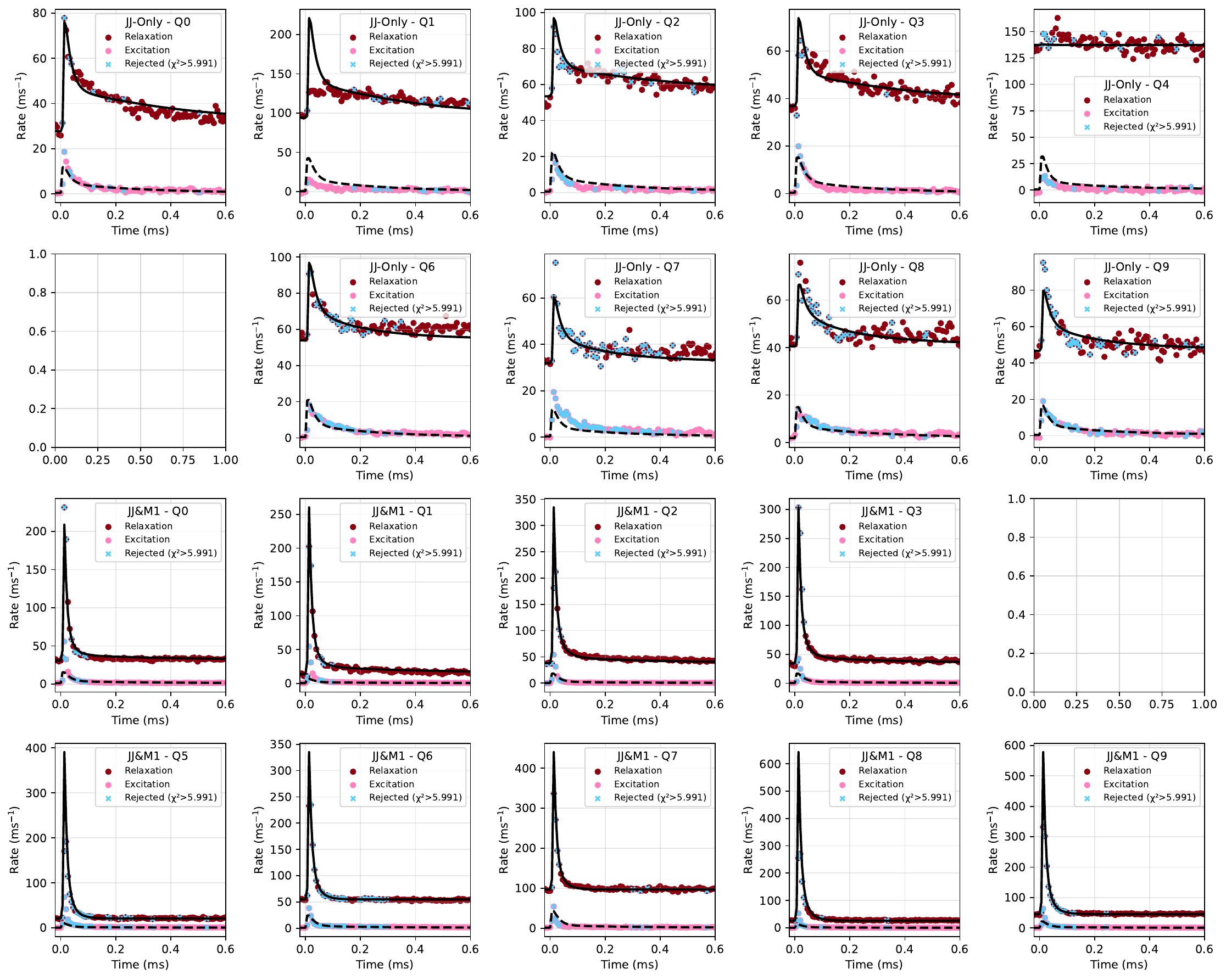}
        \caption{Comparison of the model in Sec.~\ref{sec:model_main_text} and Appendix~\ref{appendix:model_and_data} with the data on an individual qubit basis, for both relaxation (crimson) and excitation (pink) rates.  Data not passing the $\chi^2$ cut described in the main text are superimposed with blue crosses.  Model output is shown as black solid (relaxation) and dashed (excitation) lines.}
        \label{fig:modeling_by_qubit}
\end{figure*} 

\section{IQ Data}\label{appendix:iq}

\begin{figure}[tbp]
    \begin{center}
        \includegraphics[width=0.35\columnwidth]{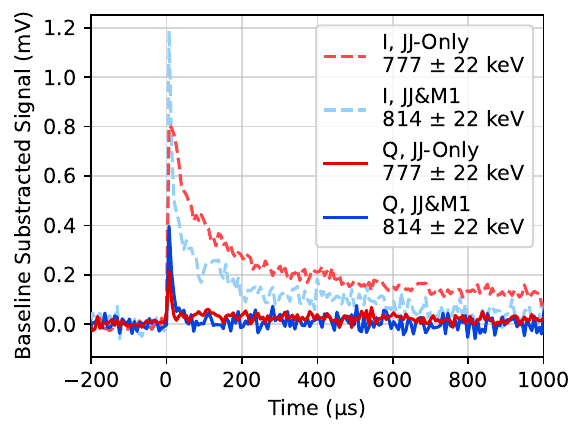}
    \end{center}
    \caption{I 
    (dashed) and Q (solid) quadrature voltages as a function of time, averaged over qubits on both chips.  The JJ-Only array is shown in red while the JJ\&M1 array is shown in blue.  Deviation in Q indicates either presence of a small $|2\rangle$ population or a shift in the readout resonator due to kinetic inductance.}
    \label{fig:iq}
\end{figure}

Figure~\ref{fig:iq} shows a deviation from the $|0\rangle-|1\rangle$ axis immediately following radiation impact.  In principle this could be due to a shift in the frequency of the readout resonator due to its kinetic inductance, or the involvement of qubit states $\geq|2\rangle$.  As the kinetic inductance of the readout resonator is insufficient to sustain a shift of this magnitude~\cite{tanakaSuperfluidStiffnessFlatBand2024}, we attribute the behavior in Fig.~\ref{fig:iq} to participation from qubit states $\geq|2\rangle$.

\end{widetext}